\documentclass[namedreferences]{solarphysics}

\usepackage[hyperref,optionalrh,showbiblabels]{spr-sola-addons} 
\usepackage{graphicx}        
\usepackage{color}           
\usepackage{breakurl}        

\newcommand{\etal}{{\it et al.}}

\newcommand{\RNum}[1]{\uppercase\expandafter{\romannumeral #1\relax}}

\chardef\us=`\_

\begin{document}

\begin{article}
\begin{opening}

\title{Photospheric Shear Flows in Solar Active Regions and Their Relation to Flare Occurrence}

\author[addressref={aff1,aff2},corref,email={shpark@isee.nagoya-u.ac.jp}]{\inits{S.-H.}\fnm{Sung-Hong}~\lnm{Park}\orcid{0000-0001-9149-6547}}
\author[addressref={aff1,aff3}]{\fnm{Jordan A.}~\lnm{Guerra}\orcid{0000-0001-8819-9648}}
\author[addressref=aff1]{\fnm{Peter T.}~\lnm{Gallagher}\orcid{0000-0001-9745-0400}}
\author[addressref=aff4]{\fnm{Manolis K. }~\lnm{Georgoulis}\orcid{0000-0001-6913-1330}}
\author[addressref=aff5]{\fnm{D. Shaun}~\lnm{Bloomfield}\orcid{0000-0002-4183-9895}}

\address[id=aff1]{School of Physics, Trinity College Dublin, College Green, Dublin 2, Ireland}
\address[id=aff2]{Institute for Space-Earth Environmental Research, Nagoya University, Nagoya, Japan}
\address[id=aff3]{Department of Physics, Villanova University, Villanova PA, USA}
\address[id=aff4]{Research Center Astronomy and Applied Mathematics, Academy of Athens, 4 Soranou Efesiou Street, 11527 Athens, Greece}
\address[id=aff5]{Northumbria University, Newcastle upon Tyne, NE1 8ST, UK}

\runningauthor{Park \etal}
\runningtitle{Active region photospheric shear flows and flares}

\begin{abstract}
Solar active regions (ARs) that produce major flares typically exhibit strong plasma shear flows around photospheric magnetic polarity inversion lines (MPILs). It is therefore important to quantitatively measure such photospheric shear flows in ARs for a better understanding of their relation to flare occurrence. Photospheric flow fields were determined by applying the Differential Affine Velocity Estimator for Vector Magnetograms (DAVE4VM) method to a large data set of 2,548 co-aligned pairs of AR vector magnetograms with 12-min separation over the period 2012--2016. From each AR flow-field map, three shear-flow parameters were derived corresponding to the mean ($<$\,$S$\,$>$), maximum ($S_\mathrm{max}$) and integral ($S_\mathrm{sum}$) shear-flow speeds along strong-gradient, strong-field MPIL segments. We calculated flaring rates within 24\,hr as a function of each shear-flow parameter, and also investigated the relation between the parameters and the waiting time ($\tau$) until the next major flare (class M1.0 or above) after the parameter observation. In general, it is found that the larger $S_\mathrm{sum}$ an AR has, the more likely it is for the AR to produce flares within 24\,hr. It is also found that among ARs which produce major flares, if one has a larger value of $S_\mathrm{sum}$ then $\tau$ generally gets shorter. These results suggest that large ARs with widespread and/or strong shear flows along MPILs tend to not only be more flare productive, but also produce major flares within 24\,hr or less.
\end{abstract}

\keywords{Active Regions, Magnetic Fields; Active Regions, Velocity Field; Flares, Relation to Magnetic Field; Velocity Fields, Photosphere}
\end{opening}

\section{Introduction}
\label{sec1}
Solar flares produce strong electromagnetic radiation, often associated with high-energy particles and coronal mass ejections (CMEs), which can disturb the state of the near-Earth space environment, known as space weather. In general, flares occur suddenly with a significant increase in brightness by a few orders of magnitude within minutes to tens of minutes. Stochasticity and abruptness in flare triggering render an accurate flare prediction a very difficult task. Over the last few decades, there have been numerous studies on active region (AR) magnetic field and plasma properties to better understand the physics underlying flare energy build-up and triggering mechanisms as well as favorable preconditions for flares \citep[\textit{e.g.}][and references therein]{2009AdSpR..43..739S,2015SoPh..290.3457S,2017NatAs...1E..85W}. Consequently, it is well known that most flares occur in solar ARs, where intense magnetic fields exist in the form of complex, non-potential structures \citep[\textit{e.g.}][]{2002SoPh..209..171G,2010ApJ...713..440J,2010ApJ...718...43P,2012SoPh..281..639L,2012ApJ...759L...4T,2016SoPh..291.1711M}.

A large number of studies on AR magnetic polarity inversion lines (MPILs) has been carried out to examine the relation between MPIL properties at single points in time and flare productivity. For example, \citet{2007ApJ...655L.117S} analyzed a data set of 2,500 randomly selected AR line-of-sight (LOS) magnetograms, calculating a parameter (log($R$)) that measures the total unsigned magnetic flux around a subset of MPILs that have strong gradients in the LOS magnetic field ($B_{\mathrm{los}}$) across them. It was found that the larger the value of log($R$), the higher the probability for major flares (M and X in the classification of the \textit{Geostationary Operational Environmental Satellites} (GOES)) to occur in the AR within 24\,hr of the measurement. \citet{2011SpWea...9.4003F} considered a parameter denoted ($\mathrm{^{L}WL_{SG}}$) as a proxy for AR free magnetic energy, defined as the integral of the strength of the horizontal gradient of $B_{\mathrm{los}}$ over all strong-field segments of MPILs for which the horizontal component of the potential magnetic field is greater than 150\,G. A positive correlation was found between $\mathrm{^{L}WL_{SG}}$ and the event rate of major flares over 24\,hr after the observation time, by considering $\sim$40,000 LOS magnetograms of $\sim$1,300 ARs.

Numerical simulations report that flaring ARs forming strong MPILs also exhibit strong plasma shear flows along them, associated mainly with emergence and/or cancellation of magnetic flux in the photosphere \citep[\textit{e.g.}][]{1999ApJ...510..485A,2003ApJ...585.1073A,2004ApJ...605L..73R,2008ASPC..383...91M}. Several observational studies also support these simulation results. \citet{2004ApJ...617L.151Y} found strong shear-flow motions along MPILs of AR NOAA 10486 prior to an X10 flare. These shear flows of persistent velocities of up to 1.6\,km\,s$^{-1}$ appeared more than 2\,hr before the flare at spatial scales of $\sim$5\,Mm around white-light flare kernels. \citet{2006ApJ...644.1278D} also observed in NOAA 10486 that there were persistent and long-lived (\textit{i.e.} $\geq$5\,hr) strong horizontal and vertical shear flows (both on the order of 1\,km\,s$^{-1}$) at MPILs until the X10 flare occurred. \citet{2009ApJ...705..821W} determined photospheric flow fields in 46 ARs by applying the Fourier Local Correlation Tracking \citep[FLCT:][]{2008ASPC..383..373F} and Differential Affine Velocity Estimator \citep[DAVE:][]{2005ApJ...632L..53S} methods to 2,708 co-aligned pairs of LOS magnetograms at 96-min time separation. Several parameters were calculated from the AR flow-field maps, such as the integral and several statistical moments of MPIL-weighted and field-weighted shear flows \citep[for details on the parameters, see Table 2 of][]{2009ApJ...705..821W}. Associating GOES flares with the magnetogram data, it was found that the shear-flow parameters are positively correlated with flare peak flux, but not as strongly as the magnetic parameter log($R$) is.

Although the investigation of AR shear-flow properties is very important to understand the triggering of flares, it has rarely been carried out with a large data set of flaring and non-flaring ARs. In this article, we examine photospheric shear flows derived from 2,548 co-aligned pairs of AR vector magnetograms at 12-min time separation over the period 2012--2016. Several parameters are determined to characterize shear flows near a subset of MPILs that exhibit strong horizontal field and strong gradients in the vertical component of magnetic field. These parameters are used to: i) study frequency distributions of ARs that do and do not produce flares in the next 24\,hr, ii) calculate probabilities of flaring in the subsequent 24\,hr as a function of each parameter, and iii) examine their relation to the waiting time to the next major flare after their measurement. This study will help to not only better understand the role of AR photospheric shear flows in relation to flare occurrence, but also potentially improve our ability to predict flares.

\section{Data and Analysis}
\label{sec2}
The \textit{Helioseismic and Magnetic Imager} \citep[HMI:][]{2012SoPh..275..207S} instrument on board the the \textit{Solar Dynamics Observatory} \citep[SDO:][]{2012SoPh..275....3P} provides a 12-min cadence data product called Space-weather HMI Active Region Patches \citep[SHARPs:][]{2014SoPh..289.3549B} that can be useful for autonomous monitoring of AR properties, as well as for flare forecasting. There are four SHARP data series, each of which includes 16 vector magnetic field parameters computed from automatically-identified HMI AR Patches (HARPs), as well as co-aligned maps of the vector and LOS magnetic field, Doppler velocity, continuum intensity and other quantities for the HARP. In this study, we use the \textit{hmi.sharp\_cea\_720s\_nrt} data series, consisting of near-realtime (NRT) SHARPs remapped from CCD coordinates to heliographic Cylindrical Equal-Area (CEA) coordinates with the three vector magnetic field components, \textit{i.e.} radial ($B_{r}$), zenithal ($B_{\theta}$) and azimuthal ($B_{\phi}$) components, recorded at a spatial sampling of 0.03\,degree\,pixel$^{-1}$. Note that SHARP CEA NRT vector magnetograms are processed through the HMI NRT pipeline with a preliminary calibration and faster azimuth disambiguation \citep[refer to][for the details of calibration procedures and the differences between the NRT and definitive data]{2014SoPh..289.3483H}.

We calculate AR photospheric flow fields using co-aligned pairs of SHARP CEA NRT vector magnetograms with time separations of 12\,min. A detailed procedure for the determination of the AR flow fields is as follows:
\begin{enumerate}
\item A large data set of SHARP CEA NRT vector magnetograms are downloaded from the Multi-Experiment Data and Operations Centre (MEDOC; \url{http://medoc.ias.u-psud.fr}) at times ($T_{\mathrm{obs}}$) closest to 06:00\,UT each day from 16 September 2012 to 13 April 2016, restricted to those SHARPs containing NOAA-numbered ARs, being within 50$^\circ$ longitude from the central meridian, and obtained before HMI new ``Mod-L'' observing scheme that uses the polarization measurements from both front and side cameras \citep{2016SPD....47.0810L}.
\item For each vector magnetogram in the data set, a vector magnetogram with the same HARP number but observed 12\,min earlier than $T_{\mathrm{obs}}$ is downloaded, if available.
\item Both images in each vector magnetogram pair are examined for off-limb pixels, with both of the corresponding images trimmed to remove them, if necessary.
\item Co-registration of image pairs is achieved via cross correlation of the $B_{r}$ images, with the resulting cross-correlation pixel offsets applied to all three magnetic field components, $B_{r}$, $B_{\theta}$, and $B_{\phi}$.
\item A plasma velocity inversion technique, called the Differential Affine Velocity Estimator for Vector Magnetograms \citep[DAVE4VM:][]{2008ApJ...683.1134S} is applied to all co-aligned vector magnetogram pairs to calculate their photospheric flow-field maps.
\end{enumerate}
Note that the apodization window size ($L$) used in DAVE4VM is set to 15\,pixels, based on a test evaluating the normal component of the magnetic induction equation over every pixel of flow-field maps calculated from randomly selected SHARP image pairs with time separation of 12\,min over different values of $L$. This is precisely the same test as in \citet{2008ApJ...683.1134S} and \citet{2012ApJ...761..105L}, but for HMI NRT SHARPs at 12-min time separation in our case. It is important to be aware that the HMI vector magnetograms have systematic errors \citep{2014SoPh..289.3483H,2016SoPh..291.1887C}, including the 12-hour periodicity and the center-to-limb variation of noise levels, respectively, due to the spacecraft orbital velocity relative to the Sun and irregular characteristics of the HMI instrument. In this study, in order to mitigate those systematic errors to some degree, we restrict the selected SHARP CEA NRT vector magnetograms to those observed at $\approx$06:00\,UT each day and with center positions within 50$^\circ$ from the central meridian. 

\begin{figure}    
\centerline{\includegraphics[width=0.88\textwidth,clip=true]{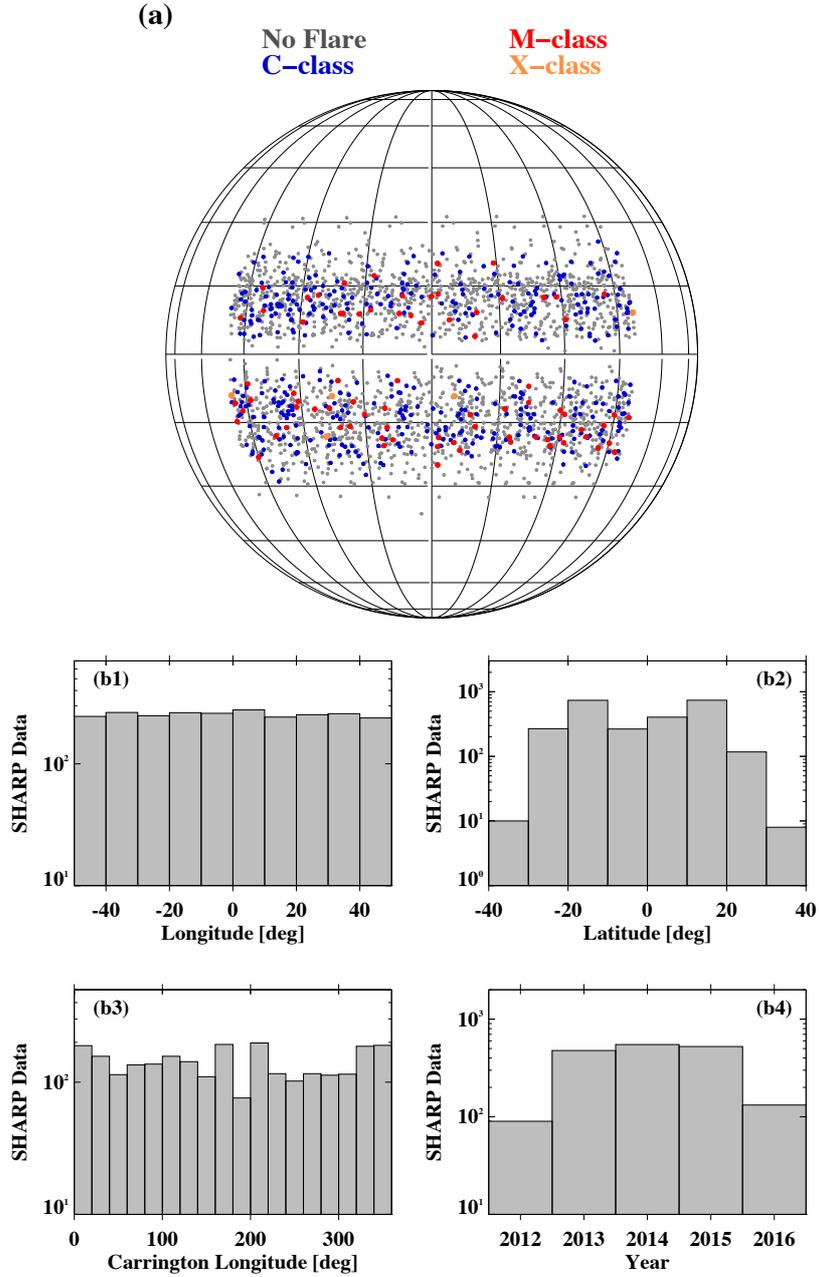}}
\caption{Properties of 2,548 SDO/HMI SHARP CEA NRT vector magnetogram pairs used to calculating AR photospheric flow fields, including (\textbf{a}) spatial locations in heliographic coordinates and (\textbf{b1}) frequency distributions of heliographic longitudes, (\textbf{b2}) heliographic latitudes, (\textbf{b3}) Carrington longitudes, and (\textbf{b4}) observation years. Heliographic coordinates in panel \textbf{a} are color-coded depending on the largest class of GOES flare assigned to the SHARP region during its entire solar disk passage: no flare as \textit{black}, C-class as \textit{blue}, M-class as \textit{red}, X-class as \textit{orange}.}
\label{f1}
\end{figure}

For each SHARP image pair, we find all the GOES soft X-rays flares that occurred in the corresponding SHARP field-of-view (FOV) during its entire passage across the solar disk. This is done searching for GOES flares with either: i) source regions the same as the NOAA AR number(s) assigned to the SHARP image pair, ii) locations falling within the FOV of the SHARP image pair. Note that we refer to the NOAA Edited Solar Event Lists (\url{ftp://ftp.swpc.noaa.gov/pub/warehouse}) for information on GOES flare start times, source regions, and locations.

A total of 2,548 co-aligned pairs of 12-min-separated SHARP CEA NRT vector magnetograms are used in this study from the time period 2012--2016. Figure~\ref{f1} shows the heliographic coordinates of the vector magnetograms at $T_{\mathrm{obs}}$ in the co-aligned pairs (panel a) as well as the distributions of heliographic longitudes (panel b1), heliographic latitudes (panel b2), Carrington longitudes (panel b3), and observed years (panel b4). The heliographic coordinates in Figure~\ref{f1}{a} are marked with different colors on the solar disk depending on the GOES class of the largest flare assigned to the vector magnetograms during the entire solar disk passage (no flare, black; C-class, blue; M-class, red; X-class, orange). Note that 522 (20\%) and 84 (3\%) of the SHARP pairs produced at least one flare above C1.0 and M1.0, respectively, within 24\,hr following observation.

\begin{figure}    
\centerline{\includegraphics[width=0.95\textwidth,clip=true]{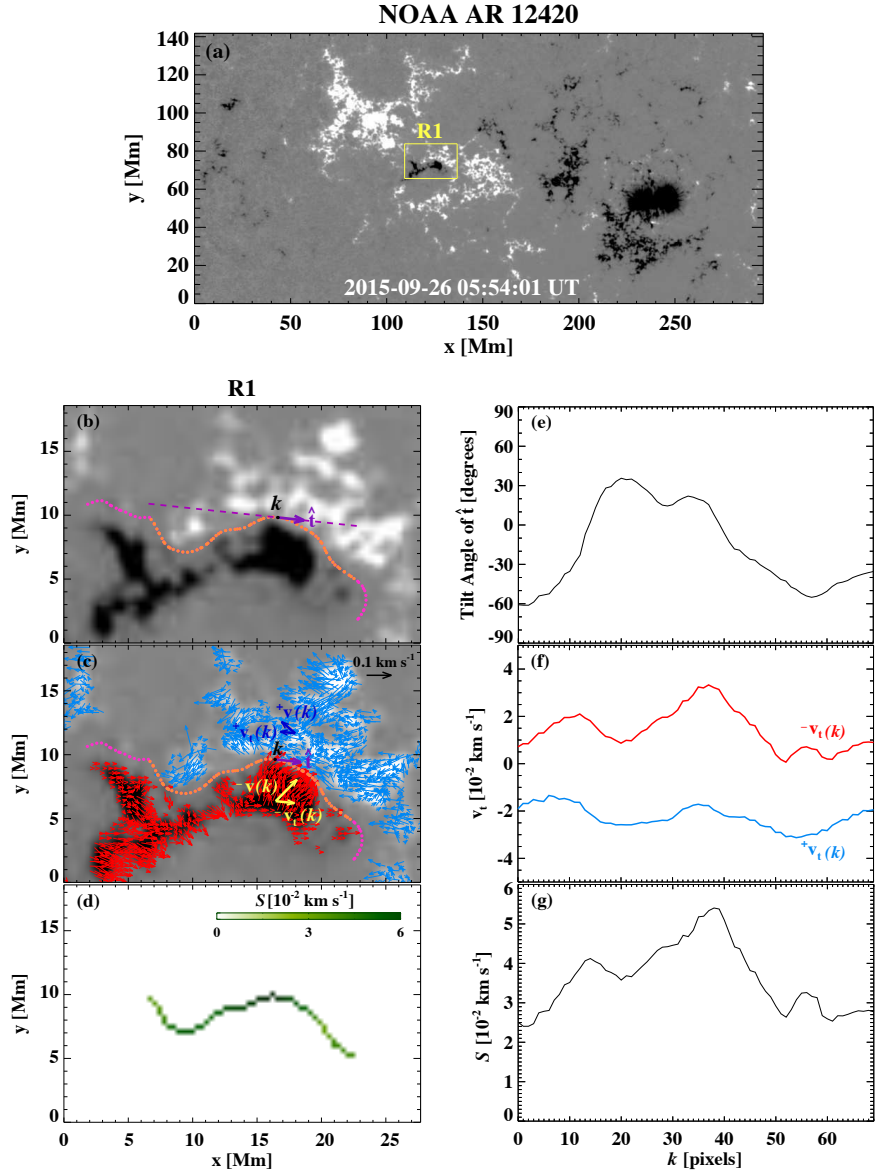}}
\caption{An example of measuring AR photospheric shear-flow speeds. \textbf{(a)} SHARP $B_{r}$ image of NOAA 12420 observed at 05:54:01\,UT on 26 September 2015 with a yellow cutout (R1) indicating the MPIL area. \textbf{(b)} MPIL (\textit{pink dotted line}) and its strong-gradient strong-field subset, $^\star$MPIL (\textit{orange dotted line}). Here $\hat{t}$ is indicated as the eastward unit vector parallel to the best-fit line (\textit{purple dashed line}) to the points of $^\star$MPIL within a local area of $1.5 \times 1.5$\,Mm centered on the $k$th pixel of $^\star$MPIL. \textbf{(c)} AR horizontal flow velocity vectors (\textit{cyan}/\textit{red arrows} on positive/negative $B_{r}$ pixels). Here $^{+}\textbf{v}\left(k\right)$ and $^{-}\textbf{v}\left(k\right)$ are the weighted mean horizontal velocity vectors from positive and negative magnetic flux pixels, respectively, within the local area of $15 \times 15$\,Mm centered on the $k$th pixel using a 2D Gaussian kernel of FWHM 4\,Mm. \textbf{(d)} AR shear-flow speed map determined as $S = |^{+}\textbf{v}_t - ^{-}\textbf{v}_t|$. \textbf{(e)} Tilt angle of $\hat{t}$ relative to the solar east, \textbf{(f)} magnitude of $^{+}\textbf{v}_t$ (\textit{cyan}, \textit{lower curve}) and $^{-}\textbf{v}_t$ (\textit{red}, \textit{upper curve}), and \textbf{(g)} $S$ as a function of pixel distance along $^\star$MPIL.}
\label{f2}
\end{figure}

Maps of shear-flow speed, $S$, are determined from each derived AR photospheric flow field, with Figure~\ref{f2} showing an example for NOAA 12420. Using the SHARP $B_{r}$ image observed at 05:54:01\,UT on 26 September 2015 (Figure~\ref{f2}{a}), a specific subset of MPILs are found, denoted $^\star$MPIL, that exhibit: i) two neighboring pixels in a smoothed $B_{r}$ image (after applying a boxcar average over $6 \times 6$\,pixel$^2$) with absolute values greater than 20\,G but different signs across an MPIL pixel; ii) horizontal magnetic field strength, $(B_{\theta}^2+B_{\phi}^2)^{1/2}$, greater than 120\,G. Note that strong-gradient, strong-field MPILs such as $^\star$MPIL have been considered the most likely places where flare-triggering flux cancellation and/or flux rope emergence can take place \citep[\textit{e.g.}][]{2010ApJ...723..634M,2012ApJ...754...15F}. In Figure~\ref{f2}{b}, an MPIL and its $^\star$MPIL subset are overlaid in pink and orange, respectively, on a cutout $B_{r}$ image of the entire NOAA 12420 (marked in Figure~\ref{f2}{a} by the yellow box R1). For the $k$th pixel on the $^\star$MPIL segment, the unit vector ($\hat{t}$) is found in the direction eastward and parallel to the best-fit line to $^\star$MPIL pixels in an area of $1.5 \times 1.5$\,Mm$^{2}$ centered on the $k$th pixel (shown in Figure~\ref{f2}{b} as a purple dashed line). Best-fit lines are found via linear least-squares regression with 2D Gaussian kernel weighting, $W_{k}$, of 4\,Mm full-width-half-maximum (FWHM) centred on the $k$th pixel. Next, two weighted mean vectors are calculated, $^{+}\textbf{v}\left(k\right)$ and $^{-}\textbf{v}\left(k\right)$, from the horizontal flow-velocity vectors of pixels with positive and negative magnetic fluxes, respectively, in an area of $15 \times 15$\,Mm$^{2}$ centered on the $k$th pixel and using the same Gaussian kernel, $W_{k}$. Figure~\ref{f2}{c} shows the horizontal velocity vectors from positive and negative magnetic fluxes as cyan and red arrows, respectively, as well as $^{+}\textbf{v}\left(k\right) $and $^{-}\textbf{v}\left(k\right)$ with blue and yellow arrows. Shear flow speed at pixel $k$ is then determined as $S\left(k\right) = |^{+}\textbf{v}_t\left(k\right) - ^{-}\textbf{v}_t\left(k\right)|$. Repeating this for all points on the $^\star$MPIL segment results in an AR shear-flow speed map as shown in Figure~\ref{f2}{d}, while the tilt angle of $\hat{t}$, the magnitude of the velocity vectors $^{+}\textbf{v}_t$ and $^{-}\textbf{v}_t$, and $S$ are shown as a function of pixel along the $^\star$MPIL in Figure~\ref{f2}{e--g}, respectively.

\section{Results}
\label{sec3}
Three shear flow parameters are calculated to examine the relation between AR photospheric flows and flare occurrence: the mean ($<$\,$S$\,$>$), maximum ($S_\mathrm{max}$), and integral ($S_\mathrm{sum}$) of $S$ along $^\star$MPILs. In addition, the mean horizontal and vertical flow speeds within $\pm$\,20\,Mm of $^\star$MPILs ($<$\,$v_{h}$\,$>$ and $<$\,$v_{z}$\,$>$, respectively) are calculated, alongside the total unsigned magnetic flux ($\Phi$) in the entire SHARP FOV for reference. Note that $<$\,$S$\,$>$, $S_\mathrm{max}$, $<$\,$v_{h}$\,$>$ and $<$\,$v_{z}$\,$>$ are \textit{intensive} parameters independent of the size of either ARs or $^\star$MPILs, while $S_\mathrm{sum}$ has an \textit{extensive} character only in case an AR possesses $^\star$MPIL. On the other hand, $\Phi$ is an extensive parameter regardless of the morphology of ARs: \textit{i.e.} the larger the AR in size, the larger the value of $\Phi$.

\begin{figure}    
\centerline{\includegraphics[width=0.96\textwidth,clip=true]{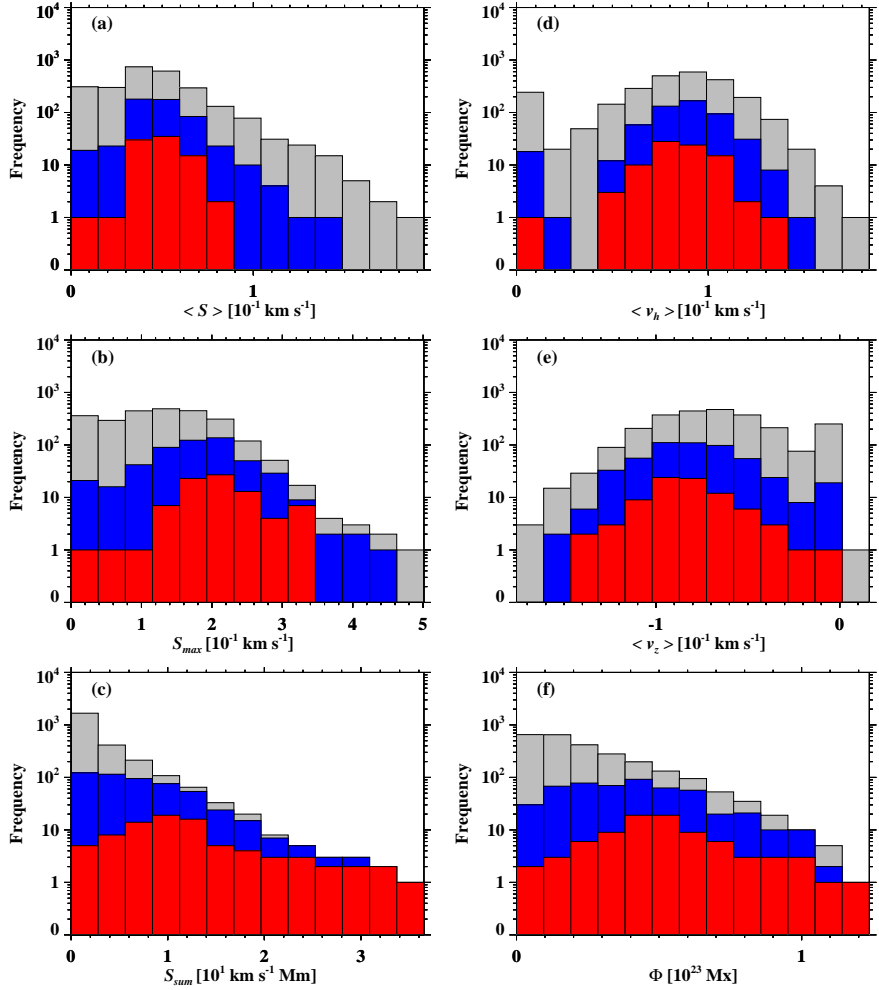}}
\caption{Frequency distributions of the six parameters under investigation. In each panel, histograms represent all SHARP image pairs (\textit{gray}) and two subsets having at least one flare within 24\,hr above C1.0 (\textit{blue}) and M1.0 (\textit{red}), respectively. (\textbf{a}--\textbf{c}) Shear flow mean $<$\,$S$\,$>$, maximum $S_\mathrm{max}$, and integral $S_\mathrm{sum}$ along strong-gradient strong-field MPIL segments $^\star$MPIL. (\textbf{d}--\textbf{e}) Mean flow speeds in the horizontal ($<$\,$v_{h}$\,$>$) and vertical ($<$\,$v_{z}$\,$>$)  directions. (\textbf{f}) Total unsigned magnetic flux ($\Phi$).}
\label{f3}
\end{figure}

Figure~\ref{f3} displays the distributions of the six parameters $<$\,$S$\,$>$, $S_\mathrm{max}$, $S_\mathrm{sum}$, $<$\,$v_{h}$\,$>$, $<$\,$v_{z}$\,$>$, and $\Phi$ for the 2,548 SHARP image pairs under investigation. The total number of entries is indicated by gray bars. The bin size for the histograms is individually selected such that all parameter ranges are represented by the same number of non-empty bins. Distributions from two subsets of the SHARP image pair data are also overplotted in each panel of Figure~\ref{f3}, namely: 1) those having at least one flare above C1.0 assigned within 24\,hr following the observation time $T_{\mathrm{obs}}$ (blue bars) and 2) those with at least one flare above M1.0 (red bars). In general, it is found that most histograms in Figure~\ref{f3} show a log-normal-like distribution with tails toward larger parameter values (\textit{i.e.} right-skewed). The distributions of the flow parameters for the entire SHARP data set show a peak or high frequency at the bin which includes zero values. This is mainly due to the fact that there is a considerably large number of SHARP image pairs containing no $^\star$MPILs, resulting in flow parameter values of zero being recorded. On the other hand, in the case of the flare-associated SHARP subsets (in particular, for the major-flaring ARs marked with red bars), we find that a peak of the histograms appears at a bin with large parameter values (\textit{i.e.} located 3--5 bins away from the zero-value bin). Similar to the shear flow parameters, $\Phi$ follows a log-normal-like, right-skewed distribution.

\begin{figure}    
\centerline{\includegraphics[width=0.96\textwidth,clip=true]{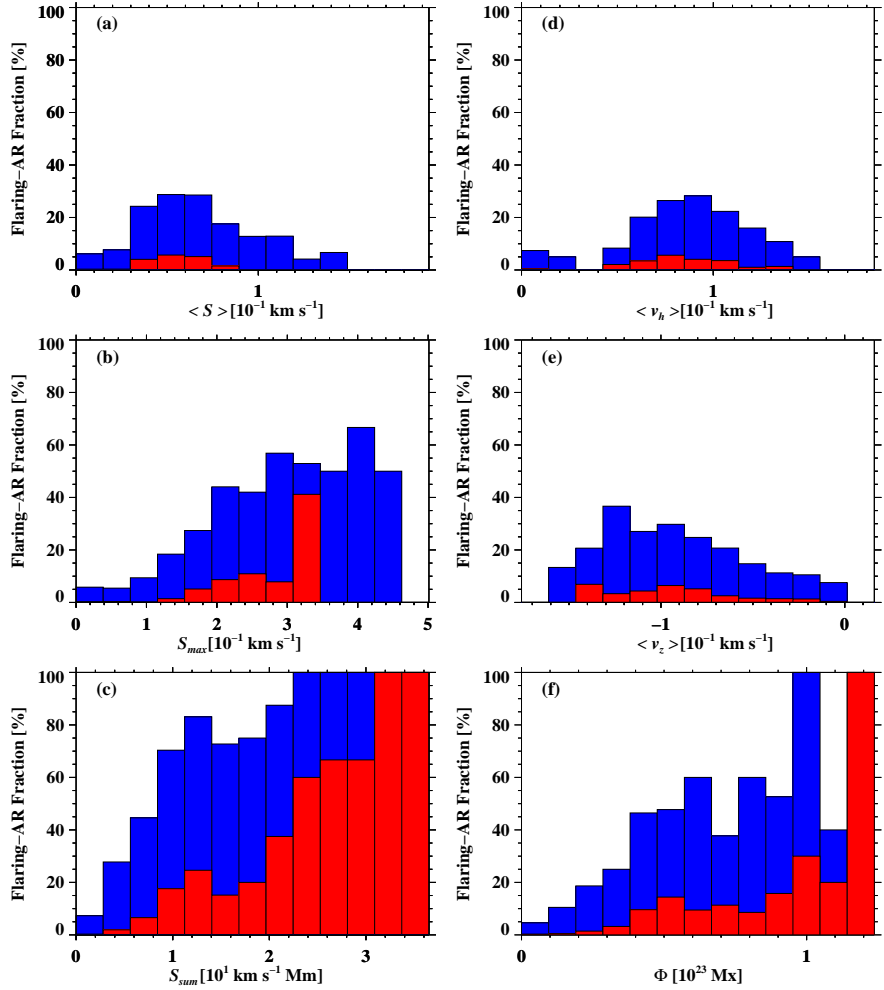}}
\caption{Flaring-AR fractions that produced at least one flare above C1.0 (\textit{blue bars}) and above M1.0 (\textit{red bars}) in the next 24\,hr, using the same panel layout as Figure~\ref{f3}.}
\label{f4}
\end{figure}

\begin{figure}    
\centerline{\includegraphics[width=0.96\textwidth,clip=true]{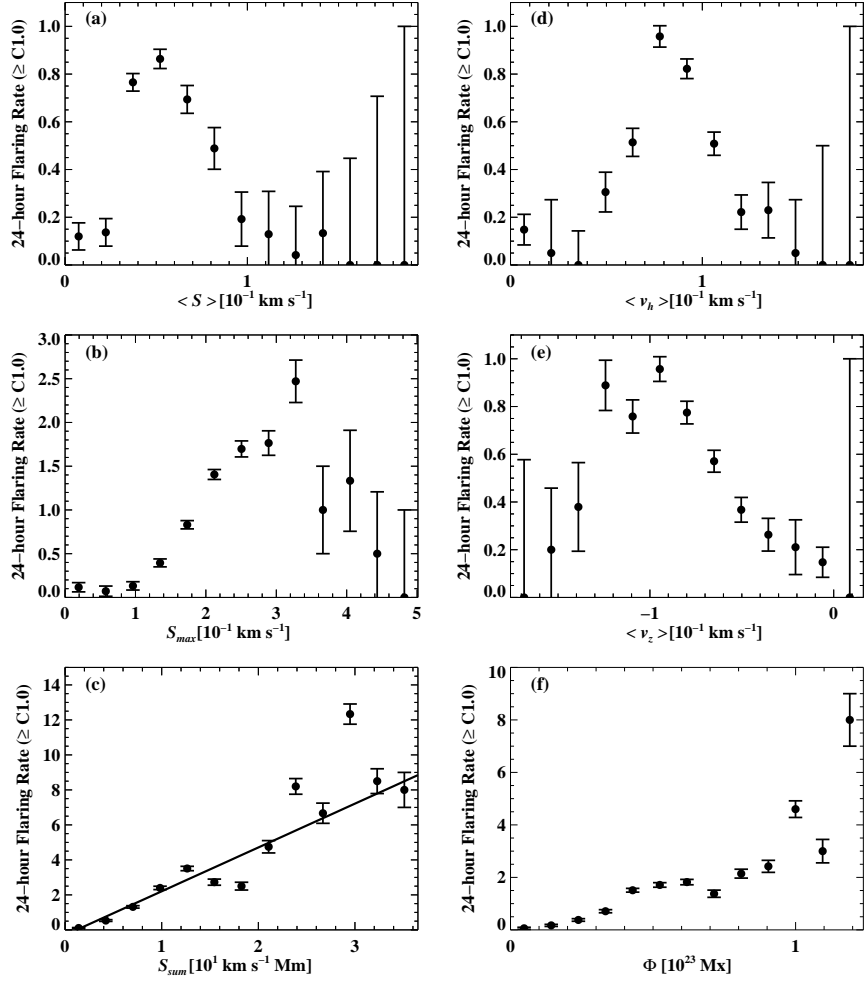}}
\caption{Flaring rates above C1.0 in the next 24\,hr as a function of each of the six parameters. Poisson uncertainties in flaring rates are depicted by error bars. In panel \textbf{c}, the \textit{solid line} indicates the best-fit linear function.}
\label{f5}
\end{figure}

\begin{figure}    
\centerline{\includegraphics[width=0.96\textwidth,clip=true]{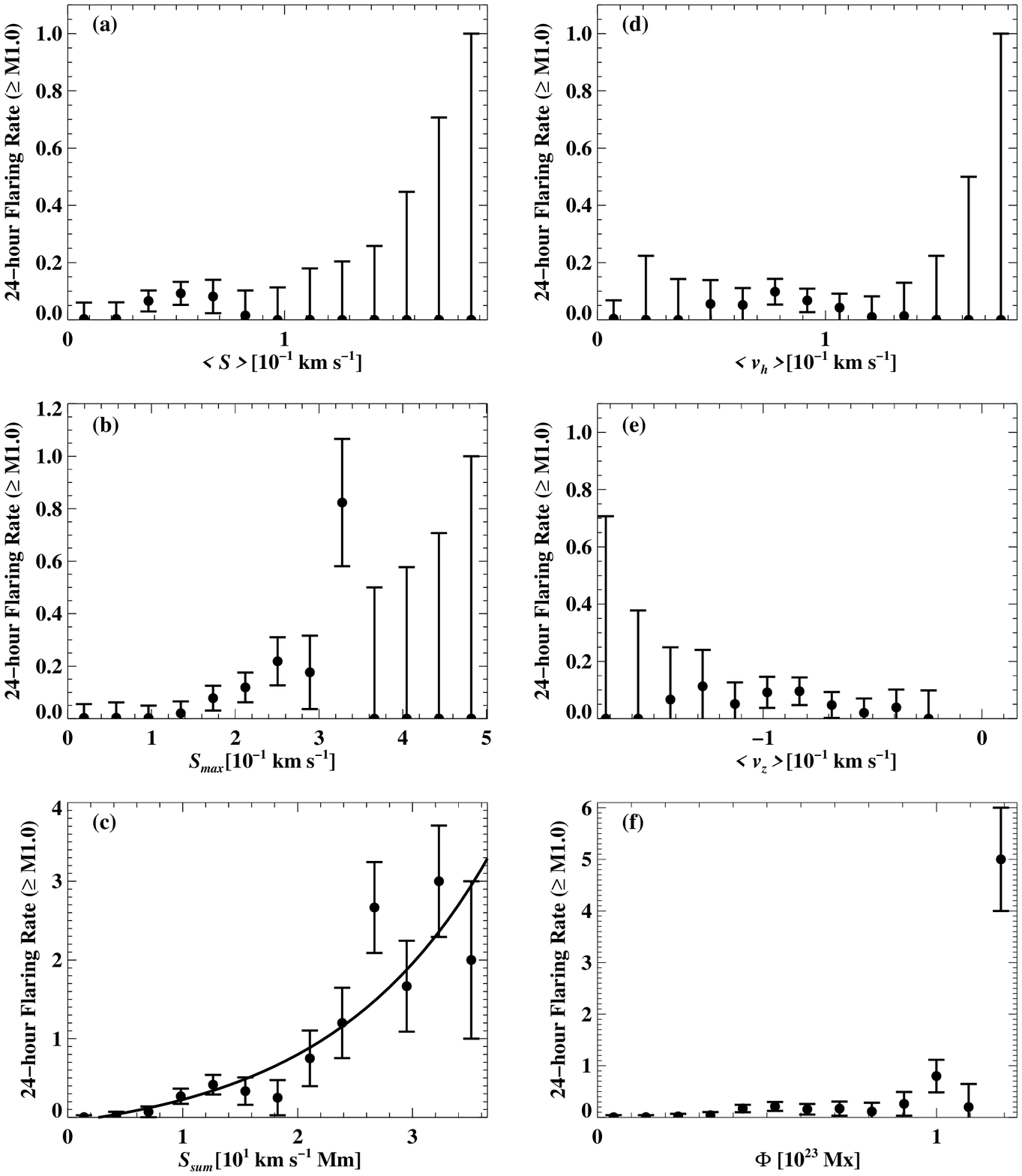}}
\caption{As Figure~\ref{f5}, but for flaring rates above M1.0 in the next 24\,hr. Poisson uncertainties in flaring rates are depicted by error bars. In panel \textbf{c}, the best-fit exponential function is marked by the \textit{solid line}.}
\label{f6}
\end{figure}

Fractions of ARs producing at least one flare in the next 24\,hr are derived from the histograms in Figure~\ref{f3}, calculated as the ratio of the number of flaring SHARP images pairs (blue or red bars for greater than C1.0 or M1.0, respectively) to the total number of SHARP images pairs in each parameter bin ($N_\mathrm{AR}$, gray bars). Figure~\ref{f4} represents these distributions of flaring-AR fractions in percentage terms for above C1.0 (blue bars) and M1.0 levels (red bars) for each of the six parameters. In general, the distributions of flaring-AR fractions can be characterized as follows:
\begin{enumerate}
\item Flaring-AR fractions for $<$\,$S$\,$>$, $<$\,$v_{h}$\,$>$, and $<$\,$v_{z}$\,$>$ show slight right skewness, with peaks of $\approx$30\% for $\geq$\,C1.0 and $\sim$5\% for $\geq$\,M1.0 at $<$\,$S$\,$> \simeq 0.05$\,km\,s$^{-1}$, $<$\,$v_{h}$\,$> \simeq 0.09$\,km\,s$^{-1}$ and $<$\,$v_{z}$\,$> \simeq -0.1$\,km\,s$^{-1}$.
\item Flaring-AR fractions with respect to $S_\mathrm{max}$ show a mostly left-skewed distribution with a peak of $\sim$70\% for $\geq$\,C1.0 and $\sim$40\% for $\geq$\,M1.0 at $S_\mathrm{max} \simeq$\,0.3--0.4\,km\,s$^{-1}$.
\item The above C1.0 flaring-AR fraction distribution for $S_\mathrm{sum}$ increases almost linearly over parameter ranges $0 \leq S_\mathrm{sum} \leq 22$\,km\,s$^{-1}$\,Mm, while above this range it shows the same 100\% flaring-AR fraction. Interestingly, over the entire range of $S_\mathrm{sum}$, flaring-AR fractions above M1.0 monotonically increase up to $\sim$100\%.
\item $\Phi$ shows a similar trend as $S_\mathrm{sum}$, but the flaring-AR fractions as function of $\Phi$ are usually smaller than those of $S_\mathrm{sum}$ over the entire parameter range with large fluctuations above $\Phi \simeq 0.6 \times 10^{23}$\,Mx.
\end{enumerate}
A remarkable finding is that the larger the value of the parameter $S_\mathrm{sum}$ an AR has, the more likely it is for the AR to produce at least one C-, M-, or X-class flare within 24\,hr. In addition, flaring-AR fractions as a function of each intensive flow parameter have relatively much smaller values compared to those of $S_\mathrm{sum}$.

\begin{figure}    
\centerline{\includegraphics[width=0.96\textwidth,clip=true]{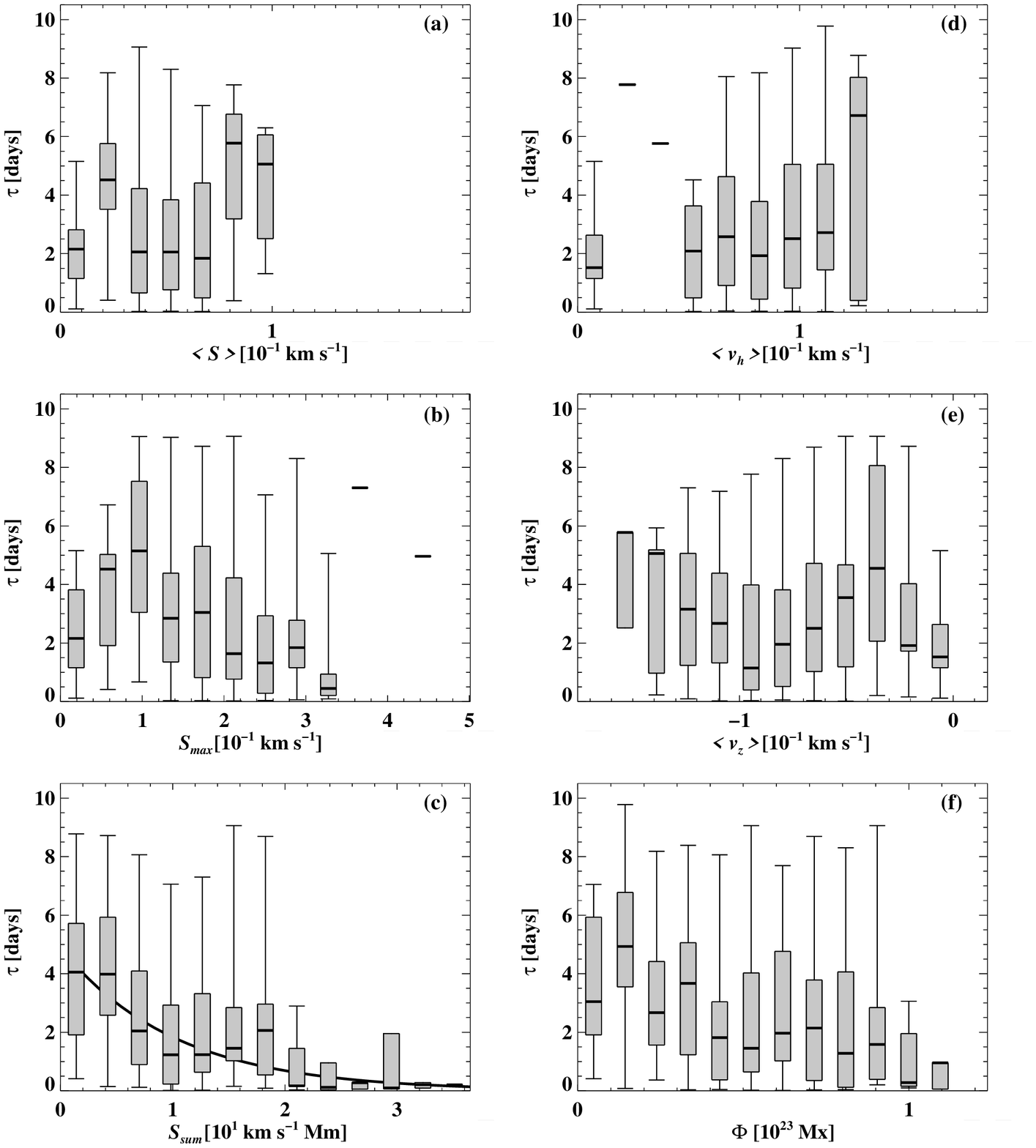}}
\caption{Box and whisker plots depicting the distribution of waiting times $\tau$ to the next major (\textit{i.e.} above M1.0) flare. A total of 1,223 SHARP image pairs for which their ARs produced at least one flare above M1.0 after the parameter measurement are classified into the same parameter bins as shown in Figures~\ref{f3}-\ref{f6}. Whiskers extend to the 2nd (lower) and 98th percentiles (upper), boxes extend from the 25th (lower edge) to the 75th percentiles (upper edge), while the 50th percentile (\textit{i.e.} median) is depicted by the horizontal bar within the box (sometimes coincident with the lower or upper edge).}
\label{f7}
\end{figure}

Flaring rates in the next 24\,hr are achieved by dividing the total numbers of flares above C1.0 and M1.0 from ARs in each parameter bin (\textit{i.e.} $N_\mathrm{\geq C1}$ and $N_\mathrm{\geq M1}$, respectively) by the number of ARs in each parameter bin $N_\mathrm{AR}$. The resulting 24-hour flaring rates, $R_\mathrm{\geq C1}$ and $R_\mathrm{\geq M1}$, are displayed in Figure~\ref{f5} and Figure~\ref{f6}, respectively. Note that the uncertainties in flaring rates indicated in these plots correspond to the $\pm$1$\sigma$ Poisson error, calculated as $1/\sqrt{N_\mathrm{AR}}$ for each parameter bin. The distributions of 24-hour flaring rates for the intensive flow parameters (panels a, b, d and e in Figures~\ref{f5} and ~\ref{f6}) as well as $\Phi$ (panel f) are very similar to those of the flaring-AR fractions shown in Figure~\ref{f4}. However, 24-hour flaring rates for $S_\mathrm{sum}$ show different characteristics to their flaring-AR fractions, \textit{i.e.} over the entire parameter range, 24-hour flaring rates above C1.0 and M1.0 show a consistently increasing trend in the form of a linear and an exponential function, respectively. Both best-fit curves, with weights of $\sqrt{N_\mathrm{AR}}$, are shown in Figures~\ref{f5}{c} and \ref{f6}{c}.

Note that panel c in Figures~\ref{f5} and \ref{f6} includes the line of best-fit for a linear and exponential function, respectively, derived from least-squares regression using weights of $\sqrt{N_\mathrm{AR}}$. From these distributions of 24-hour flaring rates, we find that: i) ARs with nearly zero-valued parameters rarely produce flares within 24\,hr of $T_{\mathrm{obs}}$, ii) ARs with increasingly larger values of $S_\mathrm{sum}$ and $\Phi$ produce increasingly more flares within 24\,hr. In the case of $R_\mathrm{\geq M1}$, this trend is more clearly shown for $S_\mathrm{sum}$, and iii) the 24-hour flaring rates above C1.0 and M1.0 as a function of $S_\mathrm{sum}$ (in unit of $10^{1}$\,km\,s$^{-1}$\,Mm) are well fit by $R_\mathrm{\geq C1} = 2.5\,S_\mathrm{sum} - 0.3$ and $R_\mathrm{\geq M1} = 0.28\exp\left(0.7\,S_\mathrm{sum} \right) - 0.3$, respectively.

The relation between point-in-time values of the six parameters and waiting time ($\tau$) until the next major flare (taken here as above M1.0) is also investigated. To achieve this, a total of 1,223 SHARP image pairs are chosen containing ARs that produced at least one flare above M1.0 after their SHARP observation times ($T_{\mathrm{obs}}$). Figure~\ref{f7} shows box and whisker plots of $\tau$ for subsets of the selected 1,223 major-flare associated SHARP image pairs in the same parameter bins used in Figures~\ref{f3}--\ref{f6}. The 2nd, 25th, 50th, 75th, and 98th percentiles of $\tau$ values in each parameter bin are denoted by the lower whisker, bottom of the box, band inside the box, top of the box, and upper whisker, respectively. In contrast to the other four parameters that appear to have no systematic dependence of $\tau$, if an AR in the major-flare associated subset has relatively large values of $S_\mathrm{sum}$ or $\Phi$, then $\tau$ for the next major flare from the AR tends to smaller time scales. However, this behavior is most clearly shown for $S_\mathrm{sum}$. For example, 81\% of ARs in the data set having $S_\mathrm{sum} \geq 20$\,km\,s$^{-1}$\,Mm produced a major flare within 24\,hr from the parameter observation time. In other words, large ARs consisting of either small $^\star$MPILs with strong shear flows or long $^\star$MPILs with at least moderate shear flows (\textit{i.e.} both leading to large $S_\mathrm{sum}$) seldom remain flare quiet, instead produce major flares on relatively short time scales. We also find that the median $\tau$ values (in unit of days) in the $S_\mathrm{sum}$ parameter bins (in unit of $10^{1}$\,km\,s$^{-1}$\,Mm) are well fit by the exponential function $\tau = 4.89 \exp\left(-0.98\,S_\mathrm{sum} \right)$. The exponential fit is marked by the solid line in Figure~\ref{f7}{c}. This suggests that $S_\mathrm{sum}$ might be practically useful for making probabilistic predictions about when the next tentative major flare will occur, as well as possibly distinguishing between flaring and flare-quiet time periods as a function of $S_\mathrm{sum}$. Note that ARs with low values of $S_\mathrm{sum}$ display a greater spread of $\tau$ values (\textit{i.e.} accessing larger values) compared to ARs with larger values of $S_\mathrm{sum}$.

\begin{figure}    
\centerline{\includegraphics[width=0.92\textwidth,clip=true]{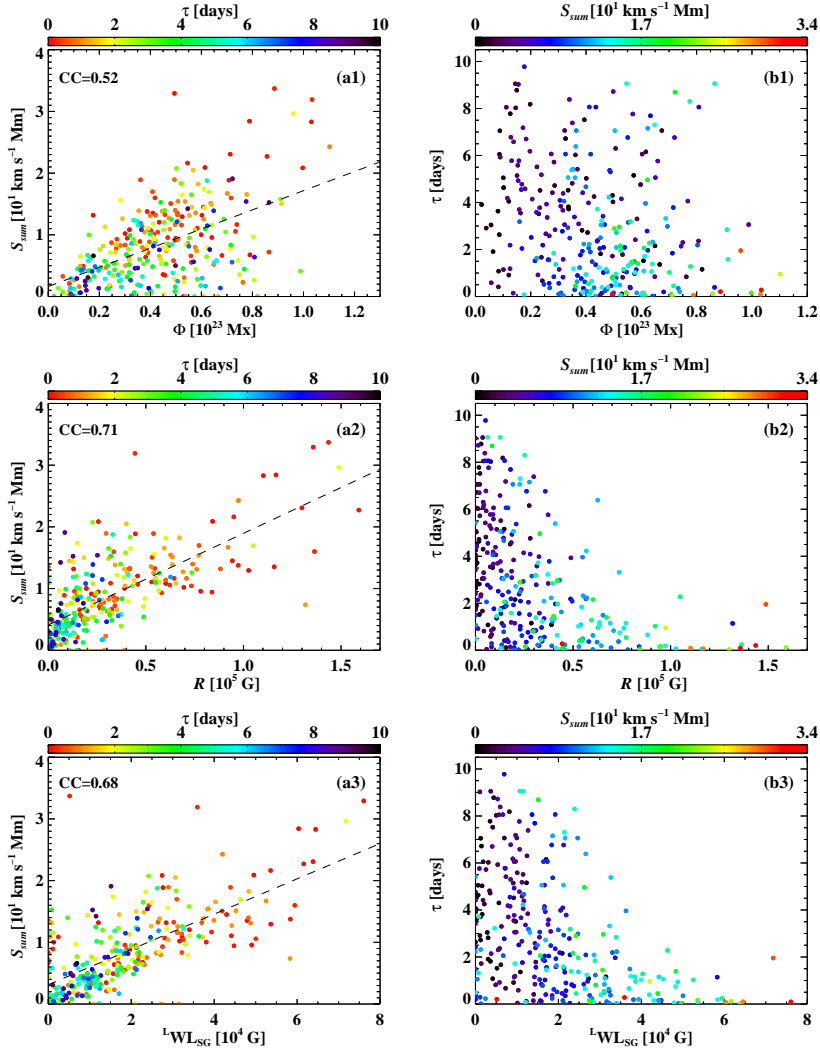}}
\caption{(\textbf{a1}) -- (\textbf{a3}) Scatter plots of $S_\mathrm{sum}$ with respect to $\Phi$, $R$ and $\mathrm{WL_{SG}}$, respectively, for a total of 296 SHARP image pairs having at least one major (\textit{i.e.} above M1.0) flare after the SHARP observation times ($T_{\mathrm{obs}}$). The Pearson correlation coefficient (CC) between $S_\mathrm{sum}$ and $\Phi$, $R$, and $\mathrm{WL_{SG}}$ is marked. The \textit{dashed} \textit{lines} indicate linear least-squares regression fits. (\textbf{b1}) -- (\textbf{b3}) Scatter plots of $\tau$ \textit{vs}. $\Phi$, $R$, and $\mathrm{WL_{SG}}$, respectively, for the same data set of the 296 SHARP pairs. In the \textit{left} and \textit{right} panels, data points are color-coded according to their $\tau$ and $S_\mathrm{sum}$ values, respectively.}
\label{f8}
\end{figure}

Here, $S_\mathrm{sum}$ presents the best flaring-potential diagnostics among the parameters under study. We examine how closely $S_\mathrm{sum}$ is correlated with $\Phi$, in order to understand whether or not these two parameters can be used as complementary parameters in relation to flare occurrence. Figure~\ref{f8} shows scatter plots of $S_\mathrm{sum}$ \textit{vs}. $\Phi$ (panel a1) and $\tau$ \textit{vs}. $\Phi$ (panel b1) for the 296 SHARP image pairs having at least one major flare (\textit{i.e.} above M1.0) after the SHARP observation times ($T_{\mathrm{obs}}$). $S_\mathrm{sum}$ and $\Phi$ are moderately correlated with each other in a positive linear-like fashion, but display a significant degree of scatter with a linear (Pearson) correlation coefficient (CC) equal to 0.54. Panel b1 of Figure~\ref{f8} shows that among ARs with values of $\Phi$ in a given range, \textit{e.g.} (0.2$-$0.8)$\times$10$^{23}$\,Mx, relatively larger values of $S_\mathrm{sum}$ (represented by data point colors) tend to correspond to shorter values of $\tau$. For a subset of ARs having values smaller than the median $\Phi$ (\textit{i.e.} 0.43$\times$10$^{23}$\,Mx) of the entire 296 ARs, it is also found that the mean of $S_\mathrm{sum}$ is 1.8 times larger in the case of ARs with $\tau$ $\mathrm{\leq}$ 24\,hr compared to the other case of those with $\tau$ $\mathrm{>}$ 24\,hr. This implies that $S_\mathrm{sum}$ and $\Phi$ are not exactly duplicated measures with respect to their relations with $\tau$, but could conceivably be jointly used for better understanding of AR flaring activity. In the same context, we examine how significantly $S_\mathrm{sum}$ is correlated with $R$ or $\mathrm{^{L}WL_{SG}}$ which are well-known, flare-prediction parameters related to the length of $^\star$MPIL, but derived from photospheric LOS magnetograms at single points in time. As shown in the panels a2 and a3 of Figure~\ref{f8}, $S_\mathrm{sum}$ shows a fairly good, positive, linear correlation with $R$ and $\mathrm{^{L}WL_{SG}}$, respectively, with CCs of 0.71 and 0.68. However, there are several data points that deviate from the linear least-squares regression lines (denoted by the dashed lines). Interestingly, ARs with very small values (less than 0.5$\times$10$^{4}$\,G) of $\mathrm{^{L}WL_{SG}}$ but with relatively large values (greater than 5\,km\,s$^{-1}$\,Mm) of $S_\mathrm{sum}$ tend to produce major flares soon (\textit{i.e.} shorter $\tau$). A similar trend is also found in the case of $R$. In addition, among ARs with values smaller than the median $\mathrm{^{L}WL_{SG}}$ (1.56$\times$10$^{4}$\,G) of the entire 296 ARs, the ones that produce a major flare in the next 24\,hr (\textit{i.e.} $\tau$ $\mathrm{\leq}$ 24\,hr) have $S_\mathrm{sum}$ 1.6 times larger on average than the others. It is therefore evident that $S_\mathrm{sum}$ can provide an additional and/or supplementary information about flare occurrence beyond what is provided by $\Phi$, $R$ or $\mathrm{^{L}WL_{SG}}$.

\section{Summary and Discussion}
\label{sec4}
In this study, AR photospheric plasma flow field maps are determined by applying the DAVE4VM method to 2,548 co-aligned pairs of SHARP vector magnetograms with time separations of 12\,min over the period 2012--2016. From each AR flow field map, the derived shear-flow parameters $<$\,$S$\,$>$, $S_\mathrm{max}$, and $S_\mathrm{sum}$ respectively represent the mean, maximum, and integral of shear-flow speeds along a strong-gradient, strong-field MPIL subset, $^\star$MPIL. In addition, the mean horizontal and vertical flow speeds around $^\star$MPILs are calculated as $<$\,$v_{h}$\,$>$ and $<$\,$v_{z}$\,$>$, respectively, as well as the total unsigned magnetic flux $\Phi$. These parameters have been systematically investigated with the large data set of the AR vector magnetograms, for the first time, to find out their relation to properties of flare occurrence such as flaring rates in the next 24\,hr and waiting time $\tau$ to next major flare (\textit{i.e.} above M1.0). As a result, it is found that:
\begin{enumerate}  
\item The larger the value of $S_\mathrm{sum}$, the more likely it is for the AR to produce flares within 24\,hr.
\item In ARs that produce at least one flare above M1.0, larger values of $S_\mathrm{sum}$ indicate that waiting times $\tau$ until the next major flare are shorter.
\item Both flaring rates and $\tau$ can be fitted by regression curves that are functions of $S_\mathrm{sum}$.
\end{enumerate}
Through this study on photospheric shear flows in ARs, we conclude that, in general, large ARs consisting of either small $^\star$MPILs with strong shear flows or long $^\star$MPILs with at least moderate shear flows seldom remain flare quiet, instead they produce major flares on short time scales. We emphasize that some of the shear flow parameters for a given AR, in particular $S_\mathrm{sum}$, could be used for flare forecasting, such as how likely the AR will be flare quiet or flare productive within 24\,hr, or tentatively when the next flare will occur if the AR has a high chance of producing a flare. For example, a flare forecasting method relying on $S_\mathrm{sum}$ could be implemented to produce flare probabilities under the assumption of Poisson statistics and using the best-fit functional forms of 24-hour flaring rates $R_\mathrm{\geq C1}$ and $R_\mathrm{\geq M1}$ reported in Section~\ref{sec3} that depend on $S_\mathrm{sum}$. It is also worthwhile to mention that the flow parameters are derived from a pair of AR magnetic field data at different points in time, which takes into account the temporal evolution of AR photospheric magnetic fields. Note that most AR properties currently used for flare prediction are derived from photospheric LOS and vector magnetograms at single points in time \citep[see, \textit{e.g.},][]{2007ApJ...656.1173L,2015ApJ...798..135B,2016ApJ...829...89B}.

Flaring-AR fractions as a function of $\Phi$ show similar, but less clear, trends to those as a function of $S_\mathrm{sum}$. Although it is found that $S_\mathrm{sum}$ and $\Phi$ are moderately correlated, they only achieve a Pearson correlation coefficient of 0.54. This implies that $S_\mathrm{sum}$ and $\Phi$ can be used as \textit{complementary} parameters in relation to flare occurrence. It has been reported that $\Phi$ is a useful baseline predictor for flare forecasting with well-known caveats, namely that it depends almost entirely on the size of ARs but is insensitive to the detailed magnetic structure of ARs \citep[see, \textit{e.g.},][]{2003ApJ...595.1296L,2012SoPh..276..161G,2016ApJ...829...89B}. Hence, any viable flare predictor must be shown to work better than $\Phi$. In this study no flare predictions are made from the parameters studied, so it will be interesting to investigate whether $S_\mathrm{sum}$ or some combination of the investigated flow parameters are more successful in flare prediction than $\Phi$.

It has been shown in the numerical magnetohydrodynamic simulations of \citet{2008ASPC..383...91M} that shear flows can be produced by the Lorentz (\textit{i.e.} tension) force in an emerging twisted magnetic flux rope. \citet{2012ApJ...761...61G} also found that strong and systematic non-neutralized currents are formed in NOAA 10930 only along $^\star$MPIL segments \citep[see also][]{2017SoPh..292..159K}. These suggest that the Lorentz force in $^\star$MPIL may be the most natural cause of shear flows and associated magnetic shear. Emerging flux, possibly evolving into flux rope formation, is thought to play  a crucial role in supplying free magnetic energy into the AR corona as an energy source for flares and coronal mass ejections \citep[\textit{e.g.}][]{2005ApJ...630..543F,2015ApJ...811...16K,2016ApJ...824...93F}, as well as in triggering these events \citep[\textit{e.g.}][]{2000ApJ...545..524C,2012ApJ...760...31K,2013ApJ...778...13P,2015ApJ...806..171Y,2016ApJ...824..148Y}. Therefore, the shear-flow parameters and their relation to flare occurrence as examined in this study may somehow be implicated with the emergence of flux ropes in ARs. This needs to be precisely studied in terms of whether shear-flow parameters pertain to characteristics of emerging flux ropes. In addition, the temporal evolution of shear-flow parameters could be of particular interest in terms of the trigger and eruption mechanism(s) of AR flux ropes. Further studies, including superposed epoch analysis of shear-flow evolution in flaring ARs, will help to more precisely understand the physics underlying flare energy build-up along with the role of AR shear flows in triggering flares and/or driving eruptive instabilities. For the time being, $S_\mathrm{sum}$ has been included as a potential flare predictor in the European Union's  Horizon 2020 Flare Likelihood And Region Eruption foreCASTing (FLARECAST) project (\url{http://flarecast.eu}).

\begin{acks}
The authors thank E. Pariat for constructive suggestions that helped clarify several topics discussed in the article. This work also benefited from discussions at the International Space Science Institute (Bern, Switzerland) International Working Team meetings on ``Improving the Reliability of Solar Eruption Predictions to Facilitate the Determination of Targets-of-Opportunity for Instruments With a Limited Field-of-View'' led by P.A. Higgins (later D.S. Bloomfield) and M.K. Georgoulis. The data used in this work are courtesy of the NASA SDO/HMI science team, as well as the GOES team. The SHARP CEA NRT vector magnetograms were provided by the MEDOC data and operations centre (CNES/CNRS/Univ. Paris-Sud; \url{http://medoc.ias.u-psud.fr}). This research has made use of NASA Astrophysics Data System (ADS). This research was funded by the European Union Horizon 2020 research and innovation programme under grant agreement No.~640216 (FLARECAST).
\end{acks}
\hfill \break
\hfill \break
\noindent \textbf{Disclosure of Potential Conflicts of Interest} The authors declare that they have no conflicts of interest.

\bibliographystyle{spr-mp-sola.bst} 
\bibliography{ref.bib}

\begin{thebibliography}{45}
\ifx\bisbn     \undefined \def\bisbn  #1{ISBN #1}\fi
\ifx\binits    \undefined \def\binits#1{#1}\fi
\ifx\bauthor   \undefined \def\bauthor#1{#1}\fi
\ifx\batitle   \undefined \def\batitle#1{#1}\fi
\ifx\bjtitle   \undefined \def\bjtitle#1{\textit{#1}}\fi
\ifx\bvolume   \undefined \def\bvolume#1{\textbf{#1}}\fi
\ifx\byear     \undefined \def\byear#1{#1}\fi
\ifx\bissue    \undefined \def\bissue#1{#1}\fi
\ifx\bfpage    \undefined \def\bfpage#1{#1}\fi
\ifx\blpage    \undefined \def\blpage #1{#1}\fi
\ifx\burl      \undefined \def\burl#1{\textsf{#1}}\fi
\ifx\href      \undefined \def\href#1#2{\textsf{#2}}\fi
\ifx\betal     \undefined \def\betal{\textit{et al.}}\fi
\ifx\bctitle   \undefined \def\bctitle#1{#1}\fi
\ifx\beditor   \undefined \def\beditor#1{#1}\fi
\ifx\bbtitle   \undefined \def\bbtitle#1{\textit{#1}}\fi
\ifx\bedition  \undefined \def\bedition#1{#1}\fi
\ifx\bseriesno \undefined \def\bseriesno#1{\textbf{#1}}\fi
\ifx\blocation \undefined \def\blocation#1{#1}\fi
\ifx\bsertitle \undefined \def\bsertitle#1{\textit{#1}}\fi
\ifx\bsnm      \undefined \def\bsnm#1{#1}\fi
\ifx\bsuffix   \undefined \def\bsuffix#1{#1}\fi
\ifx\bparticle \undefined \def\bparticle#1{#1}\fi
\ifx\barticle  \undefined \def\barticle#1{}\fi
\ifx\binstitute  \undefined \def\binstitute#1{#1}\fi
\ifx\bpublisher  \undefined \def\bpublisher#1{#1}\fi
\ifx\doiurl    \undefined
  \def\doiurl#1{\href{http://dx.doi.org/#1}{\textsf{DOI}}}\fi
\ifx\arxivurl  \undefined
  \def\arxivurl#1{\href{http://arxiv.org/abs/#1}{\textsf{arXiv}}}\fi
\ifx\adsurl    \undefined
  \def\adsurl#1{\href{http://adsabs.harvard.edu/abs/#1}{\textsf{ADS}}}\fi
\ifx\botherref \undefined \def\botherref#1{}\fi
\ifx\url       \undefined \def\url#1{\textsf{#1}}\fi
\ifx\bchapter  \undefined \def\bchapter#1{}\fi
\ifx\bbook     \undefined \def\bbook#1{}\fi
\ifx\bcomment  \undefined \def\bcomment#1{#1}\fi
\ifx\oauthor   \undefined \def\oauthor#1{#1}\fi
\ifx\citeauthoryear \undefined\def \citeauthoryear#1{#1}\fi
\ifx\endbibitem\undefined \def\endbibitem{}\fi
\ifx\bconflocation  \undefined \def\bconflocation#1{#1} \fi

\bibitem[\protect\citeauthoryear{{Amari}
  \textit{et~al.}}{2003}]{2003ApJ...585.1073A}
\begin{barticle}
\bauthor{\bsnm{{Amari}}, \binits{T.}},
\bauthor{\bsnm{{Luciani}}, \binits{J.F.}},
\bauthor{\bsnm{{Aly}}, \binits{J.J.}},
\bauthor{\bsnm{{Mikic}}, \binits{Z.}},
\bauthor{\bsnm{{Linker}}, \binits{J.}}:
\byear{2003},
\batitle{{Coronal Mass Ejection: Initiation, Magnetic Helicity, and Flux Ropes.
  I. Boundary Motion-driven Evolution}}.
\bjtitle{\apj}
\bvolume{585},
\bfpage{1073}.
\doiurl{10.1086/345501}.
\adsurl{2003ApJ...585.1073A}.
\end{barticle}
\endbibitem

\bibitem[\protect\citeauthoryear{{Antiochos}, {DeVore}, and
  {Klimchuk}}{1999}]{1999ApJ...510..485A}
\begin{barticle}
\bauthor{\bsnm{{Antiochos}}, \binits{S.K.}},
\bauthor{\bsnm{{DeVore}}, \binits{C.R.}},
\bauthor{\bsnm{{Klimchuk}}, \binits{J.A.}}:
\byear{1999},
\batitle{{A Model for Solar Coronal Mass Ejections}}.
\bjtitle{\apj}
\bvolume{510},
\bfpage{485}.
\doiurl{10.1086/306563}.
\adsurl{1999ApJ...510..485A}.
\end{barticle}
\endbibitem

\bibitem[\protect\citeauthoryear{{Barnes}
  \textit{et~al.}}{2016}]{2016ApJ...829...89B}
\begin{barticle}
\bauthor{\bsnm{{Barnes}}, \binits{G.}},
\bauthor{\bsnm{{Leka}}, \binits{K.D.}},
\bauthor{\bsnm{{Schrijver}}, \binits{C.J.}},
\bauthor{\bsnm{{Colak}}, \binits{T.}},
\bauthor{\bsnm{{Qahwaji}}, \binits{R.}},
\bauthor{\bsnm{{Ashamari}}, \binits{O.W.}},
\bauthor{\bsnm{{Yuan}}, \binits{Y.}},
\bauthor{\bsnm{{Zhang}}, \binits{J.}},
\bauthor{\bsnm{{McAteer}}, \binits{R.T.J.}},
\bauthor{\bsnm{{Bloomfield}}, \binits{D.S.}},
\bauthor{\bsnm{{Higgins}}, \binits{P.A.}},
\bauthor{\bsnm{{Gallagher}}, \binits{P.T.}},
\bauthor{\bsnm{{Falconer}}, \binits{D.A.}},
\bauthor{\bsnm{{Georgoulis}}, \binits{M.K.}},
\bauthor{\bsnm{{Wheatland}}, \binits{M.S.}},
\bauthor{\bsnm{{Balch}}, \binits{C.}},
\bauthor{\bsnm{{Dunn}}, \binits{T.}},
\bauthor{\bsnm{{Wagner}}, \binits{E.L.}}:
\byear{2016},
\batitle{{A Comparison of Flare Forecasting Methods. I. Results from the
  All-Clear Workshop}}.
\bjtitle{\apj}
\bvolume{829},
\bfpage{89}.
\doiurl{10.3847/0004-637X/829/2/89}.
\adsurl{2016ApJ...829...89B}.
\end{barticle}
\endbibitem

\bibitem[\protect\citeauthoryear{{Bobra} and
  {Couvidat}}{2015}]{2015ApJ...798..135B}
\begin{barticle}
\bauthor{\bsnm{{Bobra}}, \binits{M.G.}},
\bauthor{\bsnm{{Couvidat}}, \binits{S.}}:
\byear{2015},
\batitle{{Solar Flare Prediction Using SDO/HMI Vector Magnetic Field Data with
  a Machine-learning Algorithm}}.
\bjtitle{\apj}
\bvolume{798},
\bfpage{135}.
\doiurl{10.1088/0004-637X/798/2/135}.
\adsurl{2015ApJ...798..135B}.
\end{barticle}
\endbibitem

\bibitem[\protect\citeauthoryear{{Bobra}
  \textit{et~al.}}{2014}]{2014SoPh..289.3549B}
\begin{barticle}
\bauthor{\bsnm{{Bobra}}, \binits{M.G.}},
\bauthor{\bsnm{{Sun}}, \binits{X.}},
\bauthor{\bsnm{{Hoeksema}}, \binits{J.T.}},
\bauthor{\bsnm{{Turmon}}, \binits{M.}},
\bauthor{\bsnm{{Liu}}, \binits{Y.}},
\bauthor{\bsnm{{Hayashi}}, \binits{K.}},
\bauthor{\bsnm{{Barnes}}, \binits{G.}},
\bauthor{\bsnm{{Leka}}, \binits{K.D.}}:
\byear{2014},
\batitle{{The Helioseismic and Magnetic Imager (HMI) Vector Magnetic Field
  Pipeline: SHARPs - Space-Weather HMI Active Region Patches}}.
\bjtitle{\solphys}
\bvolume{289},
\bfpage{3549}.
\doiurl{10.1007/s11207-014-0529-3}.
\adsurl{2014SoPh..289.3549B}.
\end{barticle}
\endbibitem

\bibitem[\protect\citeauthoryear{{Chen} and
  {Shibata}}{2000}]{2000ApJ...545..524C}
\begin{barticle}
\bauthor{\bsnm{{Chen}}, \binits{P.F.}},
\bauthor{\bsnm{{Shibata}}, \binits{K.}}:
\byear{2000},
\batitle{{An Emerging Flux Trigger Mechanism for Coronal Mass Ejections}}.
\bjtitle{\apj}
\bvolume{545},
\bfpage{524}.
\doiurl{10.1086/317803}.
\adsurl{2000ApJ...545..524C}.
\end{barticle}
\endbibitem

\bibitem[\protect\citeauthoryear{{Couvidat}
  \textit{et~al.}}{2016}]{2016SoPh..291.1887C}
\begin{barticle}
\bauthor{\bsnm{{Couvidat}}, \binits{S.}},
\bauthor{\bsnm{{Schou}}, \binits{J.}},
\bauthor{\bsnm{{Hoeksema}}, \binits{J.T.}},
\bauthor{\bsnm{{Bogart}}, \binits{R.S.}},
\bauthor{\bsnm{{Bush}}, \binits{R.I.}},
\bauthor{\bsnm{{Duvall}}, \binits{T.L.}},
\bauthor{\bsnm{{Liu}}, \binits{Y.}},
\bauthor{\bsnm{{Norton}}, \binits{A.A.}},
\bauthor{\bsnm{{Scherrer}}, \binits{P.H.}}:
\byear{2016},
\batitle{{Observables Processing for the Helioseismic and Magnetic Imager
  Instrument on the Solar Dynamics Observatory}}.
\bjtitle{\solphys}
\bvolume{291},
\bfpage{1887}.
\doiurl{10.1007/s11207-016-0957-3}.
\adsurl{2016SoPh..291.1887C}.
\end{barticle}
\endbibitem

\bibitem[\protect\citeauthoryear{{Deng}
  \textit{et~al.}}{2006}]{2006ApJ...644.1278D}
\begin{barticle}
\bauthor{\bsnm{{Deng}}, \binits{N.}},
\bauthor{\bsnm{{Xu}}, \binits{Y.}},
\bauthor{\bsnm{{Yang}}, \binits{G.}},
\bauthor{\bsnm{{Cao}}, \binits{W.}},
\bauthor{\bsnm{{Liu}}, \binits{C.}},
\bauthor{\bsnm{{Rimmele}}, \binits{T.R.}},
\bauthor{\bsnm{{Wang}}, \binits{H.}},
\bauthor{\bsnm{{Denker}}, \binits{C.}}:
\byear{2006},
\batitle{{Multiwavelength Study of Flow Fields in Flaring Super Active Region
  NOAA 10486}}.
\bjtitle{\apj}
\bvolume{644},
\bfpage{1278}.
\doiurl{10.1086/503600}.
\adsurl{2006ApJ...644.1278D}.
\end{barticle}
\endbibitem

\bibitem[\protect\citeauthoryear{{Falconer}
  \textit{et~al.}}{2011}]{2011SpWea...9.4003F}
\begin{barticle}
\bauthor{\bsnm{{Falconer}}, \binits{D.}},
\bauthor{\bsnm{{Barghouty}}, \binits{A.F.}},
\bauthor{\bsnm{{Khazanov}}, \binits{I.}},
\bauthor{\bsnm{{Moore}}, \binits{R.}}:
\byear{2011},
\batitle{{A tool for empirical forecasting of major flares, coronal mass
  ejections, and solar particle events from a proxy of active-region free
  magnetic energy}}.
\bjtitle{Space Weather}
\bvolume{9},
\bfpage{S04003}.
\doiurl{10.1029/2009SW000537}.
\adsurl{2011SpWea...9.4003F}.
\end{barticle}
\endbibitem

\bibitem[\protect\citeauthoryear{{Fan}}{2005}]{2005ApJ...630..543F}
\begin{barticle}
\bauthor{\bsnm{{Fan}}, \binits{Y.}}:
\byear{2005},
\batitle{{Coronal Mass Ejections as Loss of Confinement of Kinked Magnetic Flux
  Ropes}}.
\bjtitle{\apj}
\bvolume{630},
\bfpage{543}.
\doiurl{10.1086/431733}.
\adsurl{2005ApJ...630..543F}.
\end{barticle}
\endbibitem

\bibitem[\protect\citeauthoryear{{Fan}}{2016}]{2016ApJ...824...93F}
\begin{barticle}
\bauthor{\bsnm{{Fan}}, \binits{Y.}}:
\byear{2016},
\batitle{{Modeling the Initiation of the 2006 December 13 Coronal Mass Ejection
  in AR 10930: The Structure and Dynamics of the Erupting Flux Rope}}.
\bjtitle{\apj}
\bvolume{824},
\bfpage{93}.
\doiurl{10.3847/0004-637X/824/2/93}.
\adsurl{2016ApJ...824...93F}.
\end{barticle}
\endbibitem

\bibitem[\protect\citeauthoryear{{Fang}
  \textit{et~al.}}{2012}]{2012ApJ...754...15F}
\begin{barticle}
\bauthor{\bsnm{{Fang}}, \binits{F.}},
\bauthor{\bsnm{{Manchester}}, \binits{W.} \bsuffix{IV}},
\bauthor{\bsnm{{Abbett}}, \binits{W.P.}},
\bauthor{\bsnm{{van der Holst}}, \binits{B.}}:
\byear{2012},
\batitle{{Buildup of Magnetic Shear and Free Energy during Flux Emergence and
  Cancellation}}.
\bjtitle{\apj}
\bvolume{754},
\bfpage{15}.
\doiurl{10.1088/0004-637X/754/1/15}.
\adsurl{2012ApJ...754...15F}.
\end{barticle}
\endbibitem

\bibitem[\protect\citeauthoryear{{Fisher} and
  {Welsch}}{2008}]{2008ASPC..383..373F}
\begin{bchapter}
\bauthor{\bsnm{{Fisher}}, \binits{G.H.}},
\bauthor{\bsnm{{Welsch}}, \binits{B.T.}}:
\byear{2008},
\bctitle{{FLCT: A Fast, Efficient Method for Performing Local Correlation
  Tracking}}.
In: \beditor{\bsnm{{Howe}}, \binits{R.}},
\beditor{\bsnm{{Komm}}, \binits{R.W.}},
\beditor{\bsnm{{Balasubramaniam}}, \binits{K.S.}},
\beditor{\bsnm{{Petrie}}, \binits{G.J.D.}} (eds.)
\bbtitle{Subsurface and Atmospheric Influences on Solar Activity},
\bsertitle{Astron. Soc. Pacific Conf. Ser.}
\bseriesno{383},
\bfpage{373}.
\adsurl{2008ASPC..383..373F}.
\end{bchapter}
\endbibitem

\bibitem[\protect\citeauthoryear{{Gallagher}, {Moon}, and
  {Wang}}{2002}]{2002SoPh..209..171G}
\begin{barticle}
\bauthor{\bsnm{{Gallagher}}, \binits{P.T.}},
\bauthor{\bsnm{{Moon}}, \binits{Y.-J.}},
\bauthor{\bsnm{{Wang}}, \binits{H.}}:
\byear{2002},
\batitle{{Active-Region Monitoring and Flare Forecasting I. Data Processing and
  First Results}}.
\bjtitle{\solphys}
\bvolume{209},
\bfpage{171}.
\doiurl{10.1023/A:1020950221179}.
\adsurl{2002SoPh..209..171G}.
\end{barticle}
\endbibitem

\bibitem[\protect\citeauthoryear{{Georgoulis}}{2012}]{2012SoPh..276..161G}
\begin{barticle}
\bauthor{\bsnm{{Georgoulis}}, \binits{M.K.}}:
\byear{2012},
\batitle{{Are Solar Active Regions with Major Flares More Fractal,
  Multifractal, or Turbulent Than Others?}}
\bjtitle{\solphys}
\bvolume{276},
\bfpage{161}.
\doiurl{10.1007/s11207-010-9705-2}.
\adsurl{2012SoPh..276..161G}.
\end{barticle}
\endbibitem

\bibitem[\protect\citeauthoryear{{Georgoulis}, {Titov}, and
  {Miki{\'c}}}{2012}]{2012ApJ...761...61G}
\begin{barticle}
\bauthor{\bsnm{{Georgoulis}}, \binits{M.K.}},
\bauthor{\bsnm{{Titov}}, \binits{V.S.}},
\bauthor{\bsnm{{Miki{\'c}}}, \binits{Z.}}:
\byear{2012},
\batitle{{Non-neutralized Electric Current Patterns in Solar Active Regions:
  Origin of the Shear-generating Lorentz Force}}.
\bjtitle{\apj}
\bvolume{761},
\bfpage{61}.
\doiurl{10.1088/0004-637X/761/1/61}.
\adsurl{2012ApJ...761...61G}.
\end{barticle}
\endbibitem

\bibitem[\protect\citeauthoryear{{Hoeksema}
  \textit{et~al.}}{2014}]{2014SoPh..289.3483H}
\begin{barticle}
\bauthor{\bsnm{{Hoeksema}}, \binits{J.T.}},
\bauthor{\bsnm{{Liu}}, \binits{Y.}},
\bauthor{\bsnm{{Hayashi}}, \binits{K.}},
\bauthor{\bsnm{{Sun}}, \binits{X.}},
\bauthor{\bsnm{{Schou}}, \binits{J.}},
\bauthor{\bsnm{{Couvidat}}, \binits{S.}},
\bauthor{\bsnm{{Norton}}, \binits{A.}},
\bauthor{\bsnm{{Bobra}}, \binits{M.}},
\bauthor{\bsnm{{Centeno}}, \binits{R.}},
\bauthor{\bsnm{{Leka}}, \binits{K.D.}},
\bauthor{\bsnm{{Barnes}}, \binits{G.}},
\bauthor{\bsnm{{Turmon}}, \binits{M.}}:
\byear{2014},
\batitle{{The Helioseismic and Magnetic Imager (HMI) Vector Magnetic Field
  Pipeline: Overview and Performance}}.
\bjtitle{\solphys}
\bvolume{289},
\bfpage{3483}.
\doiurl{10.1007/s11207-014-0516-8}.
\adsurl{2014SoPh..289.3483H}.
\end{barticle}
\endbibitem

\bibitem[\protect\citeauthoryear{{Jing}
  \textit{et~al.}}{2010}]{2010ApJ...713..440J}
\begin{barticle}
\bauthor{\bsnm{{Jing}}, \binits{J.}},
\bauthor{\bsnm{{Tan}}, \binits{C.}},
\bauthor{\bsnm{{Yuan}}, \binits{Y.}},
\bauthor{\bsnm{{Wang}}, \binits{B.}},
\bauthor{\bsnm{{Wiegelmann}}, \binits{T.}},
\bauthor{\bsnm{{Xu}}, \binits{Y.}},
\bauthor{\bsnm{{Wang}}, \binits{H.}}:
\byear{2010},
\batitle{{Free Magnetic Energy and Flare Productivity of Active Regions}}.
\bjtitle{\apj}
\bvolume{713},
\bfpage{440}.
\doiurl{10.1088/0004-637X/713/1/440}.
\adsurl{2010ApJ...713..440J}.
\end{barticle}
\endbibitem

\bibitem[\protect\citeauthoryear{{Kazachenko}
  \textit{et~al.}}{2015}]{2015ApJ...811...16K}
\begin{barticle}
\bauthor{\bsnm{{Kazachenko}}, \binits{M.D.}},
\bauthor{\bsnm{{Fisher}}, \binits{G.H.}},
\bauthor{\bsnm{{Welsch}}, \binits{B.T.}},
\bauthor{\bsnm{{Liu}}, \binits{Y.}},
\bauthor{\bsnm{{Sun}}, \binits{X.}}:
\byear{2015},
\batitle{{Photospheric Electric Fields and Energy Fluxes in the Eruptive Active
  Region NOAA 11158}}.
\bjtitle{\apj}
\bvolume{811},
\bfpage{16}.
\doiurl{10.1088/0004-637X/811/1/16}.
\adsurl{2015ApJ...811...16K}.
\end{barticle}
\endbibitem

\bibitem[\protect\citeauthoryear{{Kontogiannis}
  \textit{et~al.}}{2017}]{2017SoPh..292..159K}
\begin{barticle}
\bauthor{\bsnm{{Kontogiannis}}, \binits{I.}},
\bauthor{\bsnm{{Georgoulis}}, \binits{M.K.}},
\bauthor{\bsnm{{Park}}, \binits{S.-H.}},
\bauthor{\bsnm{{Guerra}}, \binits{J.A.}}:
\byear{2017},
\batitle{{Non-neutralized Electric Currents in Solar Active Regions and Flare
  Productivity}}.
\bjtitle{\solphys}
\bvolume{292},
\bfpage{159}.
\doiurl{10.1007/s11207-017-1185-1}.
\adsurl{2017SoPh..292..159K}.
\end{barticle}
\endbibitem

\bibitem[\protect\citeauthoryear{{Kusano}
  \textit{et~al.}}{2012}]{2012ApJ...760...31K}
\begin{barticle}
\bauthor{\bsnm{{Kusano}}, \binits{K.}},
\bauthor{\bsnm{{Bamba}}, \binits{Y.}},
\bauthor{\bsnm{{Yamamoto}}, \binits{T.T.}},
\bauthor{\bsnm{{Iida}}, \binits{Y.}},
\bauthor{\bsnm{{Toriumi}}, \binits{S.}},
\bauthor{\bsnm{{Asai}}, \binits{A.}}:
\byear{2012},
\batitle{{Magnetic Field Structures Triggering Solar Flares and Coronal Mass
  Ejections}}.
\bjtitle{\apj}
\bvolume{760},
\bfpage{31}.
\doiurl{10.1088/0004-637X/760/1/31}.
\adsurl{2012ApJ...760...31K}.
\end{barticle}
\endbibitem

\bibitem[\protect\citeauthoryear{{Lee}
  \textit{et~al.}}{2012}]{2012SoPh..281..639L}
\begin{barticle}
\bauthor{\bsnm{{Lee}}, \binits{K.}},
\bauthor{\bsnm{{Moon}}, \binits{Y.-J.}},
\bauthor{\bsnm{{Lee}}, \binits{J.-Y.}},
\bauthor{\bsnm{{Lee}}, \binits{K.-S.}},
\bauthor{\bsnm{{Na}}, \binits{H.}}:
\byear{2012},
\batitle{{Solar Flare Occurrence Rate and Probability in Terms of the Sunspot
  Classification Supplemented with Sunspot Area and Its Changes}}.
\bjtitle{\solphys}
\bvolume{281},
\bfpage{639}.
\doiurl{10.1007/s11207-012-0091-9}.
\adsurl{2012SoPh..281..639L}.
\end{barticle}
\endbibitem

\bibitem[\protect\citeauthoryear{{Leka} and
  {Barnes}}{2003}]{2003ApJ...595.1296L}
\begin{barticle}
\bauthor{\bsnm{{Leka}}, \binits{K.D.}},
\bauthor{\bsnm{{Barnes}}, \binits{G.}}:
\byear{2003},
\batitle{{Photospheric Magnetic Field Properties of Flaring versus Flare-quiet
  Active Regions. II. Discriminant Analysis}}.
\bjtitle{\apj}
\bvolume{595},
\bfpage{1296}.
\doiurl{10.1086/377512}.
\adsurl{2003ApJ...595.1296L}.
\end{barticle}
\endbibitem

\bibitem[\protect\citeauthoryear{{Leka} and
  {Barnes}}{2007}]{2007ApJ...656.1173L}
\begin{barticle}
\bauthor{\bsnm{{Leka}}, \binits{K.D.}},
\bauthor{\bsnm{{Barnes}}, \binits{G.}}:
\byear{2007},
\batitle{{Photospheric Magnetic Field Properties of Flaring versus Flare-quiet
  Active Regions. IV. A Statistically Significant Sample}}.
\bjtitle{\apj}
\bvolume{656},
\bfpage{1173}.
\doiurl{10.1086/510282}.
\adsurl{2007ApJ...656.1173L}.
\end{barticle}
\endbibitem

\bibitem[\protect\citeauthoryear{{Liu} and
  {Schuck}}{2012}]{2012ApJ...761..105L}
\begin{barticle}
\bauthor{\bsnm{{Liu}}, \binits{Y.}},
\bauthor{\bsnm{{Schuck}}, \binits{P.W.}}:
\byear{2012},
\batitle{{Magnetic Energy and Helicity in Two Emerging Active Regions in the
  Sun}}.
\bjtitle{\apj}
\bvolume{761},
\bfpage{105}.
\doiurl{10.1088/0004-637X/761/2/105}.
\adsurl{2012ApJ...761..105L}.
\end{barticle}
\endbibitem

\bibitem[\protect\citeauthoryear{{Liu}
  \textit{et~al.}}{2016}]{2016SPD....47.0810L}
\begin{barticle}
\bauthor{\bsnm{{Liu}}, \binits{Y.}},
\bauthor{\bsnm{{Baldner}}, \binits{C.}},
\bauthor{\bsnm{{Bogart}}, \binits{R.S.}},
\bauthor{\bsnm{{Bush}}, \binits{R.}},
\bauthor{\bsnm{{Couvidat}}, \binits{S.}},
\bauthor{\bsnm{{Duvall}}, \binits{T.L.}},
\bauthor{\bsnm{{Hoeksema}}, \binits{J.T.}},
\bauthor{\bsnm{{Norton}}, \binits{A.A.}},
\bauthor{\bsnm{{Scherrer}}, \binits{P.H.}},
\bauthor{\bsnm{{Schou}}, \binits{J.}}:
\byear{2016},
\batitle{{On HMI's Mod-L Sequence: Test and Evaluation}}.
\bjtitle{Amer. Astron. Soc./Solar Phys. Div. Abs.}
\bvolume{47},
\bfpage{8.10}.
\adsurl{2016SPD....47.0810L}.
\end{barticle}
\endbibitem

\bibitem[\protect\citeauthoryear{{Manchester}}{2008}]{2008ASPC..383...91M}
\begin{bchapter}
\bauthor{\bsnm{{Manchester}}, \binits{W.}}:
\byear{2008},
\bctitle{{Shear Flows Driven by the Lorentz Force: An Energy Source for Coronal
  Mass Ejections and Flares}}.
In: \beditor{\bsnm{{Howe}}, \binits{R.}},
\beditor{\bsnm{{Komm}}, \binits{R.W.}},
\beditor{\bsnm{{Balasubramaniam}}, \binits{K.S.}},
\beditor{\bsnm{{Petrie}}, \binits{G.J.D.}} (eds.)
\bbtitle{Subsurface and Atmospheric Influences on Solar Activity},
\bsertitle{Astron. Soc. Pacific Conf. Ser.}
\bseriesno{383},
\bfpage{91}.
\adsurl{2008ASPC..383...91M}.
\end{bchapter}
\endbibitem

\bibitem[\protect\citeauthoryear{{Mason} and
  {Hoeksema}}{2010}]{2010ApJ...723..634M}
\begin{barticle}
\bauthor{\bsnm{{Mason}}, \binits{J.P.}},
\bauthor{\bsnm{{Hoeksema}}, \binits{J.T.}}:
\byear{2010},
\batitle{{Testing Automated Solar Flare Forecasting with 13 Years of Michelson
  Doppler Imager Magnetograms}}.
\bjtitle{\apj}
\bvolume{723},
\bfpage{634}.
\doiurl{10.1088/0004-637X/723/1/634}.
\adsurl{2010ApJ...723..634M}.
\end{barticle}
\endbibitem

\bibitem[\protect\citeauthoryear{{McCloskey}, {Gallagher}, and
  {Bloomfield}}{2016}]{2016SoPh..291.1711M}
\begin{barticle}
\bauthor{\bsnm{{McCloskey}}, \binits{A.E.}},
\bauthor{\bsnm{{Gallagher}}, \binits{P.T.}},
\bauthor{\bsnm{{Bloomfield}}, \binits{D.S.}}:
\byear{2016},
\batitle{{Flaring Rates and the Evolution of Sunspot Group McIntosh
  Classifications}}.
\bjtitle{\solphys}
\bvolume{291},
\bfpage{1711}.
\doiurl{10.1007/s11207-016-0933-y}.
\adsurl{2016SoPh..291.1711M}.
\end{barticle}
\endbibitem

\bibitem[\protect\citeauthoryear{{Park}, {Chae}, and
  {Wang}}{2010}]{2010ApJ...718...43P}
\begin{barticle}
\bauthor{\bsnm{{Park}}, \binits{S.-h.}},
\bauthor{\bsnm{{Chae}}, \binits{J.}},
\bauthor{\bsnm{{Wang}}, \binits{H.}}:
\byear{2010},
\batitle{{Productivity of Solar Flares and Magnetic Helicity Injection in
  Active Regions}}.
\bjtitle{\apj}
\bvolume{718},
\bfpage{43}.
\doiurl{10.1088/0004-637X/718/1/43}.
\adsurl{2010ApJ...718...43P}.
\end{barticle}
\endbibitem

\bibitem[\protect\citeauthoryear{{Park}
  \textit{et~al.}}{2013}]{2013ApJ...778...13P}
\begin{barticle}
\bauthor{\bsnm{{Park}}, \binits{S.-H.}},
\bauthor{\bsnm{{Kusano}}, \binits{K.}},
\bauthor{\bsnm{{Cho}}, \binits{K.-S.}},
\bauthor{\bsnm{{Chae}}, \binits{J.}},
\bauthor{\bsnm{{Bong}}, \binits{S.-C.}},
\bauthor{\bsnm{{Kumar}}, \binits{P.}},
\bauthor{\bsnm{{Park}}, \binits{S.-Y.}},
\bauthor{\bsnm{{Kim}}, \binits{Y.-H.}},
\bauthor{\bsnm{{Park}}, \binits{Y.-D.}}:
\byear{2013},
\batitle{{Study of Magnetic Helicity Injection in the Active Region NOAA 9236
  Producing Multiple Flare-associated Coronal Mass Ejection Events}}.
\bjtitle{\apj}
\bvolume{778},
\bfpage{13}.
\doiurl{10.1088/0004-637X/778/1/13}.
\adsurl{2013ApJ...778...13P}.
\end{barticle}
\endbibitem

\bibitem[\protect\citeauthoryear{{Pesnell}, {Thompson}, and
  {Chamberlin}}{2012}]{2012SoPh..275....3P}
\begin{barticle}
\bauthor{\bsnm{{Pesnell}}, \binits{W.D.}},
\bauthor{\bsnm{{Thompson}}, \binits{B.J.}},
\bauthor{\bsnm{{Chamberlin}}, \binits{P.C.}}:
\byear{2012},
\batitle{{The Solar Dynamics Observatory (SDO)}}.
\bjtitle{\solphys}
\bvolume{275},
\bfpage{3}.
\doiurl{10.1007/s11207-011-9841-3}.
\adsurl{2012SoPh..275....3P}.
\end{barticle}
\endbibitem

\bibitem[\protect\citeauthoryear{{Roussev}
  \textit{et~al.}}{2004}]{2004ApJ...605L..73R}
\begin{barticle}
\bauthor{\bsnm{{Roussev}}, \binits{I.I.}},
\bauthor{\bsnm{{Sokolov}}, \binits{I.V.}},
\bauthor{\bsnm{{Forbes}}, \binits{T.G.}},
\bauthor{\bsnm{{Gombosi}}, \binits{T.I.}},
\bauthor{\bsnm{{Lee}}, \binits{M.A.}},
\bauthor{\bsnm{{Sakai}}, \binits{J.I.}}:
\byear{2004},
\batitle{{A Numerical Model of a Coronal Mass Ejection: Shock Development with
  Implications for the Acceleration of GeV Protons}}.
\bjtitle{\apjl}
\bvolume{605},
\bfpage{L73}.
\doiurl{10.1086/392504}.
\adsurl{2004ApJ...605L..73R}.
\end{barticle}
\endbibitem

\bibitem[\protect\citeauthoryear{{Scherrer}
  \textit{et~al.}}{2012}]{2012SoPh..275..207S}
\begin{barticle}
\bauthor{\bsnm{{Scherrer}}, \binits{P.H.}},
\bauthor{\bsnm{{Schou}}, \binits{J.}},
\bauthor{\bsnm{{Bush}}, \binits{R.I.}},
\bauthor{\bsnm{{Kosovichev}}, \binits{A.G.}},
\bauthor{\bsnm{{Bogart}}, \binits{R.S.}},
\bauthor{\bsnm{{Hoeksema}}, \binits{J.T.}},
\bauthor{\bsnm{{Liu}}, \binits{Y.}},
\bauthor{\bsnm{{Duvall}}, \binits{T.L.}},
\bauthor{\bsnm{{Zhao}}, \binits{J.}},
\bauthor{\bsnm{{Title}}, \binits{A.M.}},
\bauthor{\bsnm{{Schrijver}}, \binits{C.J.}},
\bauthor{\bsnm{{Tarbell}}, \binits{T.D.}},
\bauthor{\bsnm{{Tomczyk}}, \binits{S.}}:
\byear{2012},
\batitle{{The Helioseismic and Magnetic Imager (HMI) Investigation for the
  Solar Dynamics Observatory (SDO)}}.
\bjtitle{\solphys}
\bvolume{275},
\bfpage{207}.
\doiurl{10.1007/s11207-011-9834-2}.
\adsurl{2012SoPh..275..207S}.
\end{barticle}
\endbibitem

\bibitem[\protect\citeauthoryear{{Schmieder}, {Aulanier}, and {Vr{\v
  s}nak}}{2015}]{2015SoPh..290.3457S}
\begin{barticle}
\bauthor{\bsnm{{Schmieder}}, \binits{B.}},
\bauthor{\bsnm{{Aulanier}}, \binits{G.}},
\bauthor{\bsnm{{Vr{\v s}nak}}, \binits{B.}}:
\byear{2015},
\batitle{{Flare-CME Models: An Observational Perspective (Invited Review)}}.
\bjtitle{\solphys}
\bvolume{290},
\bfpage{3457}.
\doiurl{10.1007/s11207-015-0712-1}.
\adsurl{2015SoPh..290.3457S}.
\end{barticle}
\endbibitem

\bibitem[\protect\citeauthoryear{{Schrijver}}{2007}]{2007ApJ...655L.117S}
\begin{barticle}
\bauthor{\bsnm{{Schrijver}}, \binits{C.J.}}:
\byear{2007},
\batitle{{A Characteristic Magnetic Field Pattern Associated with All Major
  Solar Flares and Its Use in Flare Forecasting}}.
\bjtitle{\apjl}
\bvolume{655},
\bfpage{L117}.
\doiurl{10.1086/511857}.
\adsurl{2007ApJ...655L.117S}.
\end{barticle}
\endbibitem

\bibitem[\protect\citeauthoryear{{Schrijver}}{2009}]{2009AdSpR..43..739S}
\begin{barticle}
\bauthor{\bsnm{{Schrijver}}, \binits{C.J.}}:
\byear{2009},
\batitle{{Driving major solar flares and eruptions: A review}}.
\bjtitle{Adv. in Space Res.}
\bvolume{43},
\bfpage{739}.
\doiurl{10.1016/j.asr.2008.11.004}.
\adsurl{2009AdSpR..43..739S}.
\end{barticle}
\endbibitem

\bibitem[\protect\citeauthoryear{{Schuck}}{2005}]{2005ApJ...632L..53S}
\begin{barticle}
\bauthor{\bsnm{{Schuck}}, \binits{P.W.}}:
\byear{2005},
\batitle{{Local Correlation Tracking and the Magnetic Induction Equation}}.
\bjtitle{\apjl}
\bvolume{632},
\bfpage{L53}.
\doiurl{10.1086/497633}.
\adsurl{2005ApJ...632L..53S}.
\end{barticle}
\endbibitem

\bibitem[\protect\citeauthoryear{{Schuck}}{2008}]{2008ApJ...683.1134S}
\begin{barticle}
\bauthor{\bsnm{{Schuck}}, \binits{P.W.}}:
\byear{2008},
\batitle{{Tracking Vector Magnetograms with the Magnetic Induction Equation}}.
\bjtitle{\apj}
\bvolume{683},
\bfpage{1134}.
\doiurl{10.1086/589434}.
\adsurl{2008ApJ...683.1134S}.
\end{barticle}
\endbibitem

\bibitem[\protect\citeauthoryear{{Tziotziou}, {Georgoulis}, and
  {Raouafi}}{2012}]{2012ApJ...759L...4T}
\begin{barticle}
\bauthor{\bsnm{{Tziotziou}}, \binits{K.}},
\bauthor{\bsnm{{Georgoulis}}, \binits{M.K.}},
\bauthor{\bsnm{{Raouafi}}, \binits{N.-E.}}:
\byear{2012},
\batitle{{The Magnetic Energy-Helicity Diagram of Solar Active Regions}}.
\bjtitle{\apjl}
\bvolume{759},
\bfpage{L4}.
\doiurl{10.1088/2041-8205/759/1/L4}.
\adsurl{2012ApJ...759L...4T}.
\end{barticle}
\endbibitem

\bibitem[\protect\citeauthoryear{{Wang}
  \textit{et~al.}}{2017}]{2017NatAs...1E..85W}
\begin{barticle}
\bauthor{\bsnm{{Wang}}, \binits{H.}},
\bauthor{\bsnm{{Liu}}, \binits{C.}},
\bauthor{\bsnm{{Ahn}}, \binits{K.}},
\bauthor{\bsnm{{Xu}}, \binits{Y.}},
\bauthor{\bsnm{{Jing}}, \binits{J.}},
\bauthor{\bsnm{{Deng}}, \binits{N.}},
\bauthor{\bsnm{{Huang}}, \binits{N.}},
\bauthor{\bsnm{{Liu}}, \binits{R.}},
\bauthor{\bsnm{{Kusano}}, \binits{K.}},
\bauthor{\bsnm{{Fleishman}}, \binits{G.D.}},
\bauthor{\bsnm{{Gary}}, \binits{D.E.}},
\bauthor{\bsnm{{Cao}}, \binits{W.}}:
\byear{2017},
\batitle{{High-resolution observations of flare precursors in the low solar
  atmosphere}}.
\bjtitle{Nature Astronomy}
\bvolume{1},
\bfpage{0085}.
\doiurl{10.1038/s41550-017-0085}.
\adsurl{2017NatAs...1E..85W}.
\end{barticle}
\endbibitem

\bibitem[\protect\citeauthoryear{{Welsch}
  \textit{et~al.}}{2009}]{2009ApJ...705..821W}
\begin{barticle}
\bauthor{\bsnm{{Welsch}}, \binits{B.T.}},
\bauthor{\bsnm{{Li}}, \binits{Y.}},
\bauthor{\bsnm{{Schuck}}, \binits{P.W.}},
\bauthor{\bsnm{{Fisher}}, \binits{G.H.}}:
\byear{2009},
\batitle{{What is the Relationship Between Photospheric Flow Fields and Solar
  Flares?}}
\bjtitle{\apj}
\bvolume{705},
\bfpage{821}.
\doiurl{10.1088/0004-637X/705/1/821}.
\adsurl{2009ApJ...705..821W}.
\end{barticle}
\endbibitem

\bibitem[\protect\citeauthoryear{{Yang}
  \textit{et~al.}}{2004}]{2004ApJ...617L.151Y}
\begin{barticle}
\bauthor{\bsnm{{Yang}}, \binits{G.}},
\bauthor{\bsnm{{Xu}}, \binits{Y.}},
\bauthor{\bsnm{{Cao}}, \binits{W.}},
\bauthor{\bsnm{{Wang}}, \binits{H.}},
\bauthor{\bsnm{{Denker}}, \binits{C.}},
\bauthor{\bsnm{{Rimmele}}, \binits{T.R.}}:
\byear{2004},
\batitle{{Photospheric Shear Flows along the Magnetic Neutral Line of Active
  Region 10486 prior to an X10 Flare}}.
\bjtitle{\apjl}
\bvolume{617},
\bfpage{L151}.
\doiurl{10.1086/427210}.
\adsurl{2004ApJ...617L.151Y}.
\end{barticle}
\endbibitem

\bibitem[\protect\citeauthoryear{{Yang}, {Guo}, and
  {Ding}}{2015}]{2015ApJ...806..171Y}
\begin{barticle}
\bauthor{\bsnm{{Yang}}, \binits{K.}},
\bauthor{\bsnm{{Guo}}, \binits{Y.}},
\bauthor{\bsnm{{Ding}}, \binits{M.D.}}:
\byear{2015},
\batitle{{On the 2012 October 23 Circular Ribbon Flare: Emission Features and
  Magnetic Topology}}.
\bjtitle{\apj}
\bvolume{806},
\bfpage{171}.
\doiurl{10.1088/0004-637X/806/2/171}.
\adsurl{2015ApJ...806..171Y}.
\end{barticle}
\endbibitem

\bibitem[\protect\citeauthoryear{{Yang}, {Guo}, and
  {Ding}}{2016}]{2016ApJ...824..148Y}
\begin{barticle}
\bauthor{\bsnm{{Yang}}, \binits{K.}},
\bauthor{\bsnm{{Guo}}, \binits{Y.}},
\bauthor{\bsnm{{Ding}}, \binits{M.D.}}:
\byear{2016},
\batitle{{Quantifying the Topology and Evolution of a Magnetic Flux Rope
  Associated with Multi-flare Activities}}.
\bjtitle{\apj}
\bvolume{824},
\bfpage{148}.
\doiurl{10.3847/0004-637X/824/2/148}.
\adsurl{2016ApJ...824..148Y}.
\end{barticle}
\endbibitem

\end{thebibliography}

\end{article} 

\end{document}